\documentclass[journal,a4paper]{IEEEtran}
\usepackage[utf8]{inputenc}
\usepackage[english]{babel}
\usepackage[T1]{fontenc}
\usepackage{amsmath,amsfonts,amssymb,amsthm}
\usepackage{mathtools}
\usepackage{cite}
\usepackage{graphicx}
\usepackage{color}
\usepackage{epstopdf}
\usepackage{adjustbox}
\usepackage{booktabs}
\usepackage{my_tikz}
\usepackage{subfig}
\usepackage{multirow}

\def\ve#1{{\mathchoice{\mbox{\boldmath$\displaystyle #1$}}%
		{\mbox{\boldmath$\textstyle #1$}}%
		{\mbox{\boldmath$\scriptstyle #1$}}%
		{\mbox{\boldmath$\scriptscriptstyle #1$}}}}
\def\argmax{\mathop{\mathrm{argmax}}}
\def\argmin{\mathop{\mathrm{argmin}}}
\def\argmin{\mathop{\mathrm{argmin}}}

\def\isequiv{\mathrel{\lower2pt\hbox{$\widehat{\raise2pt\hbox{$=$}}$}}}

\def\A{\mathcal{A}}

\def\G{\mathbb{G}}
\def\Z{\mathbb{Z}}
\DeclareMathSymbol{\F}{\mathalpha}{AMSb}{"46}

\def\j{\mathrm{j}}

\def\herm{\mathsf{H}}
\def\E{\mathrm{E}}

\def\EsT{E_{\mathrm{s,Tx}}}

\def\No{N_{{0}}}

\def\max{\mathop{\mathrm{max}}}
\newcommand{\tr}[1]{\ensuremath{\mathrm{trace}\left(#1\right)}}

\def\NA{N_\mathrm{A}}
\def\NB{N_\mathrm{B}}
\def\NE{N_\mathrm{E}}

\def\fc{f_\mathrm{c}}
\def\co{\mathrm{c}_0}
\def \I{\ve{I}}

\def\Lambdab{\ve{\Lambda}_\mathrm{b}}
\def\Lambdaa{\ve{\Lambda}_\mathrm{a}}
\def\modM{\mathrm{mod}_{\Lambdab}}

\def\Hbul{\ve{H}_{\bullet\mathrm{A}}}
\def\HE{\ve{H}_\mathrm{EA}}
\def\HB{\ve{H}_\mathrm{BA}}
\def \HbulT{\ve{H}_{\bullet\mathrm{A,F}}}
\def \HET{\ve{H}_\mathrm{EA,F}}
\def \HBT{\ve{H}_\mathrm{BA,F}}
\def \HAug{\ve{\mathcal{H}}_\mathrm{BA}}

\def \FLPE{\ve{F}}
\def \Fo{\ve{F}^\prime}

\def \Gbul{\ve{G}_\bullet}
\def \GB{\ve{G}_\mathrm{B}}
\def \GE{\ve{G}_\mathrm{E}}

\def \Cs{C_\mathrm{S}}

\def \Rs{R_\mathrm{S}}
\def \Rbul{R_{\bullet\mathrm{A}}}
\def \RB{R_\mathrm{BA}}
\def \RE{R_\mathrm{EA}}

\def \W{\ve{W}}
\def \ZLRA{\ve{Z}}

\def \x{\ve{x}}
\def \v{\ve{v}}
\def \a{\ve{a}}

\def \nbul{\ve{n}_\bullet}
\def \nB{\ve{n}_\mathrm{B}}
\def \nE{\ve{n}_\mathrm{E}}

\def \neffbul{\ve{\tilde{n}}_\bullet}

\def \aVar{\sigma_a^2}
\def \nVar{\sigma_n^2}
\def \neffbulVar{\sigma_{\tilde{n}_\bullet}^2}
\def \neffBVar{\sigma_{\tilde{n}_\mathrm{B}}^2}
\def \neffEVar{\sigma_{\tilde{n}_\mathrm{E}}^2}

\def \ybul{\ve{y}_\bullet}
\def \yB{\ve{y}_\mathrm{B}}
\def \yE{\ve{y}_\mathrm{E}}

\def \rB{\ve{r}_\mathrm{B}}
\def \rE{\ve{r}_\mathrm{E}}

\def\CsPE{LP-$\Rs$} 
\def\ZCsPE{LRP-$\Rs$} 

\def\LRP{LRP} 
\def\LP{LP}
\def\LRAPE{LRP-$E_\mathrm{B}$} 
\def\LPE{LP-$E_\mathrm{B}$}
\def\LPC{LP-$R_\mathrm{B}$}
\def\LRPC{LRP-$R_\mathrm{B}$}
\def\LPSER{LP-$E_\mathrm{S}$}
\def\LRPSER{LRP-$E_\mathrm{S}$}

\def\LE{LE} 
\def\SD{SD} 

\newcommand{\unit}[1]{\ensuremath{\,\mathrm{#1}}}
\newcommand{\inlinemark}[2]{{
		\centering
\begin{tikzpicture}[trim left=0, trim right=1]
	\begin{axis}[
		hide axis, 
		scale only axis, 
		width=2ex, 
		height=2ex, 
		xmin=0, xmax=1, ymin=0, ymax=1
		]
		
		\addplot[
		only marks, 
		mark=#1, 
		mark size=0.8ex, 
		color=#2,
		] coordinates {(0.5,0.5)};
	\end{axis}
\end{tikzpicture}
}}

\begin{document}
\IEEEoverridecommandlockouts
\title{Preprocessing for Physical-Layer Security
	in Wireless THz-Communication}
\author{Rebekka Schulz, Robert F.H. Fischer, \IEEEmembership{Senior Member, IEEE}

\thanks{This work was supported by the
	German Federal Ministry of Education and Research (BMBF)
	under Grant 16KIS1243.}%
\thanks{Rebekka Schulz and Robert F.H. Fischer are with the
        Institute of Communications Engineering, Ulm University,
        89081 Ulm, Germany
        (e-mail: rebekka.schulz@uni-ulm.de, robert.fischer@uni-ulm.de).
}
}

\maketitle
%
\begin{abstract}
	In this paper, the usage of preprocessing to achieve physical-layer security in a wireless THz-MIMO scenario is investigated. The goal is a reliable and secure communication. Optimization of the preprocessing is done either based on the error performance or the transmission rate. For both criteria, we present a variant that is based only on the legitimate receiver or also includes the eavesdropper. For each variant, linear and lattice-reduction-aided approaches are considered. Numerical simulations are used to assess the resulting secrecy rates and error ratios. A comparison between all variants is compiled and the possible trade-offs are discussed.
\end{abstract}

%
%

\section{Introduction}
	\label{sec_introduction}
	\noindent
	Security is an important aspect of modern communication systems. For many different scenarios, low-complexity algorithms are needed to ensure secure communication. We consider wireless THz-communication using multiple antennas at transmitter and receiver. Communication in the THz-band allows flexible, high-speed communication using large bandwidths \cite{rappaport2019WirelessCommunicationsApplications, giordani20206GNetworksUse}. Due to the small size of the equipment, it can be used in easily adaptable industrial scenarios.
	
	We investigate the usage of preprocessing for physical-layer security (PLS) in an industrial wiretap scenario. The transmitter, denoted Alice, communicates with the legitimate receiver, called Bob. Additionally to ensuring reliability, security against Eve, a passive eavesdropper, has to be guaranteed. The usage of preprocessing allows for low-complexity processing at the legitimate receiver while exploiting the differences in the physical transmission channel to the eavesdropper. 
	
	In the literature usually the secrecy capacity is considered \cite{oggier2011SecrecyCapacityMIMO,khisti2010SecureTransmissionMultiple}. Recently, research has focused on the design of low-complexity algorithms to find the secrecy capacity and the optimal input covariance matrix \cite{mukherjee2021SecrecyCapacityMIMO, li2013AlternatingOptimizationAlgorithm, loyka2015AlgorithmGlobalMaximization,nguyen2020LowComplexityAlgorithmAchieving, zhao2015RobustBeamformingDesign}. In \cite{schraml2021MultiuserMIMOConcept}, preprocessing for PLS is investigated for satellite communication. However, for the calculations, knowledge about the channel to the eavesdropper has to be obtained by the transmitter, which is not feasible in reality. In \cite{mukherjee2011RobustBeamformingSecurity}, artificial interference is used, for which no knowledge about the eavesdropper is necessary. In addition to evaluating the secrecy rate, a common metric is to consider the resulting error ratios to evaluate the achieved security via the security gap \cite{klinc2011LDPCCodesGaussian, mukherjee2009UtilityBeamformingStrategies}.
	
	Preprocessing for PLS is commonly done based on the secrecy rate and requires knowledge about the eavesdropper. On the other hand classical linear preequalization aims at minimizing the error ratio of the legitimate receiver without taking the security aspect or the eavesdropper into account \cite{fischer2002PrecodingSignalShaping}. In this work, we compare both approaches and additionally consider two novel variants. One is based on the rate of the channel from Alice to Bob and is thus independent of the eavesdropper. The other is based on the resulting error ratios of both Bob and Eve. All variants are compared from an information-theoretic viewpoint, by assessing the secrecy rates, and regarding the practical performance by considering the error ratios.
	
	It was shown in \cite{schulz2023LatticeReductionAidedPreequalizationPhysicalLayer} that lattice-reduction-aided (LRA) preequalization is beneficial for PLS. It allows an improvement in performance for the legitimate receiver while at the same time preventing the eavesdropper from gaining necessary channel knowledge. Here, we use the LRA approach that was proposed in \cite{schulz2024LatticeReductionAidedPreprocessingPhysicalLayer}, which may be used for each of the employed optimization criteria. The resulting preprocessing variants are compared regarding the improvement of performance and security.

	The paper is structured as follows. In Sec.~\ref{sec_systemModel}, the channel model is presented and the communication scenario is introduced. Subsequently, the residual equalization that is performed by the receivers is presented. The employed optimization criteria are explained in Sec.~\ref{sec_preprocessing}. Additionally, the LRA approaches are introduced. In Sec.~\ref{sec_numericalResults}, numerical results of the simulations are presented. Finally, a summary and conclusions are given in Sec.~\ref{sec_conclusion}.
	
%
%

\section{System Model and Security Metric}
	\label{sec_systemModel}
\begin{figure}[b]
		\centerline{
\begin{tikzpicture}
\newcommand{\combineXY}[2]{#1 |- #2}
\node (aliceleft) at (-0.2,0.08){};
\node (alicemiddle) at (0,0.08){};
\node (aliceright) at (0.2,0.08){};

\node (bobleft) at (-0.1,-1.4){};
\node (bobright) at (0.1,-1.4) {};

\node (eveleft) at (1.8,-1.4){};
\node (everight) at (2.0,-1.4) {};

\fill[fill=gray!30](aliceleft.center) --(-1,-1.5) -- (1,-1.5) -- (aliceright.center);


\draw (aliceleft.center) -- +(0,0.1) (aliceleft.center)-- +(-0.08,-0.1) (aliceleft.center) -- +(0.08,-0.1);
\draw (alicemiddle.center) -- +(0,0.1) (alicemiddle.center)-- +(-0.08,-0.1) (alicemiddle.center) -- +(0.08,-0.1);
\draw (aliceright.center) -- +(0,0.1) (aliceright.center)-- +(-0.08,-0.1) (aliceright.center) -- +(0.08,-0.1);
\draw[white] (aliceleft.north west) -- (aliceright.north east) node[midway, above, black] {Alice};

\draw (bobleft.center) -- +(0,-0.1)  (bobleft.center)-- +(-0.08,0.1) (bobleft.center) -- +(0.08,0.1);
\draw (bobright.center) -- +(0,-0.1)  (bobright.center)-- +(-0.08,0.1) (bobright.center) -- +(0.08,0.1);
\draw[white] (bobleft.south west) -- (bobright.south east) node[midway, below, black] {Bob};


\draw (eveleft.center) -- +(0.08,-0.1)  (eveleft.center)-- +(-0.14,0.04) (eveleft.center) -- +(0.04,0.14);
\draw (everight.center) -- +(0.08,-0.1)  (everight.center)-- +(-0.14,0.04) (everight.center) -- +(0.04,0.14);
\draw[white] (eveleft.south west) -- (everight.south east) node[midway, below, black] {Eve};

\draw ([xshift=-30]\combineXY{aliceleft.west}{bobleft.south}) -- ([xshift=60]\combineXY{aliceright.east}{bobleft.south});
\end{tikzpicture}}
		\caption{\label{fig_setting}
			Considered transmission scenario.
	}
	\end{figure}
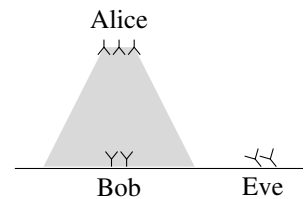

	\noindent
	We consider an industrial indoor setting, where THz-communication is used in a wireless MIMO transmission scheme. As depicted in Fig.~\ref{fig_setting}, the transmitter Alice is located on the ceiling and communicates with the legitimate receiver Bob, which corresponds to a machine below the transmitter.
	Additionally, an eavesdropper Eve is present, who tries to recover the message communicated from Alice to Bob, which should be kept secret as it might contain corporate secrets about manufacturing processes. This communication scenario corresponds to a wiretap channel \cite{wyner1975WireTapChannel}.  We assume that the presence of the eavesdropper in the direct vicinity of Bob can be prevented, thus Eve's location has a larger distance. 
	\subsection{Channel Model}\label{subsec_channelModel}\noindent
		We employ a simple channel model (cf. also \cite{schulz2023LatticeReductionAidedPreequalizationPhysicalLayer}) to model the single-carrier MIMO communication in the THz-band, which corresponds to a carrier frequency $\fc$ above 100\unit{GHz}. The used bandwidth is denoted by $B$. For this frequency band, large reflection and scattering losses occur, hence only line-of-sight transmission is considered \cite{han2014MultiRayChannelModeling,piesiewicz2007ScatteringAnalysisModeling}. The attenuation is described by the free-space path loss and can be calculated according to \cite{friis1946NoteSimpleTransmission}
		\begin{align}
			H_{\mathrm{fspl}}(d) &= \left(\frac{\co}{4\pi\fc d}\right)^2 \;,
		\end{align}
		where $\fc$ is the carrier frequency, $\co$ is the speed of light, and $d$
		denotes the distance between transmit and receive antennas. 
		
		We consider communication using multiple antennas, where Alice, Bob, and Eve use $\NA, \NB$, and $\NE$ antennas, respectively. Hereby, $\NA \ge \NB$.
		The end-to-end transfer function also includes the respective transmit and receive antenna gains and the resulting phase shift from the propagation. The transmission from transmit antenna $i=1,\dots,\NA$, to receive antenna $j = 1, \dots, \NB$ of Bob or $j = 1,\dots, \NE$ of Eve, respectively, is given by
		\cite{friis1946NoteSimpleTransmission}
		\begin{align}					      \label{eq_TransferFct}
			H_{j,i}^{\bullet\mathrm{A}} &= G_\mathrm{A} G_{\bullet} \, H_{\mathrm{fspl}} (d^{\bullet\mathrm{A}}_{j,i}) \exp \left( -\j \pi \frac{\fc}{\co} d^{\bullet\mathrm{A}}_{j,i} \right) \;,
		\end{align}
		where $\bullet \in \{\mathrm{B},\mathrm{E}\}$. Hereby, $G_\mathrm{A}$, $G_\mathrm{B}$, and $G_\mathrm{E}$ are the respective antenna gains for Alice, Bob, and Eve, dependent on the angles of arrival and departure, and $d^{\mathrm{BA}}_{j,i}$ and
		$d^{\mathrm{EA}}_{j,i}$ are the distances between the respective antennas.
		
		The $\NB\times\NA$ channel matrix $\HB$ includes the channel coefficients that describe the channel from Alice to Bob, while the $\NE\times\NA$ channel matrix $\HE$ corresponds to the channel from Alice to Eve.

		The receivers experience i.i.d. additive white Gaussian thermal noise with noise power \cite{anderson2006DigitalTransmissionEngineering}
		\begin{align}							\label{eq_Noise}
			\nVar = \mathrm{k}_\mathrm{B}TBF \;,
		\end{align}
		where $\mathrm{k}_\mathrm{B} = 1.38 \cdot 10^{-23}\unit{\frac{J}{K}}$
		is the Boltzmann constant, $T$ is the temperature in Kelvin, and $F$ is the noise factor of the receiver front-end.
		We assume that Bob and Eve use the same receiver front-end and thus experience the same noise variance $\nVar$. We consider the transmitted energy per symbol in relation to the noise power spectral density, which is calculated according to $\EsT/\No= P/\nVar$, where $P$ is the sum transmit power.
	
	\subsection{Communication Scenario} \label{sec_scenario}
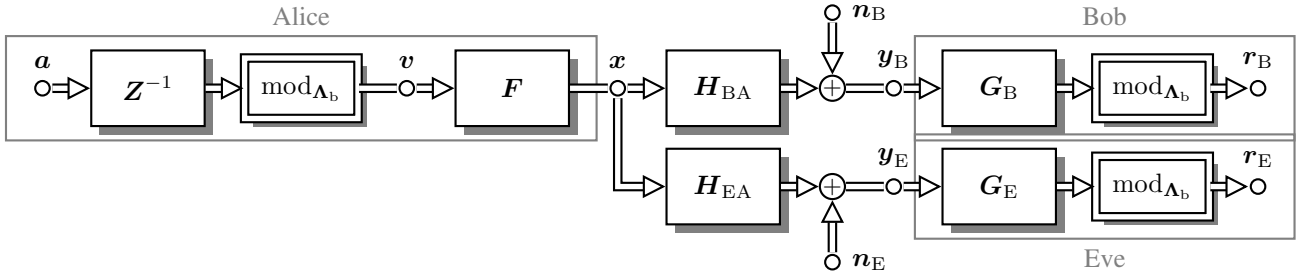
\begin{figure*}\centering

\begin{tikzpicture}[auto, thick, node distance=\blockDist, >={Triangle[open,length=3mm,width=3mm]}, double distance = 2pt]
\draw
node at (0,0)[label={above:$\a$},inoutput] (input1) {\phantom{.}}
node [filter, right= of input1] (ZTx) {$\ZLRA^{-1}$}
node [filter, right=of ZTx, minimum height = 0.8cm, double] (modRedTx) {$\modM$}
node [label={$\v$},inoutput, right =of modRedTx](v){\phantom{.}}
node [filter, right=of v] (FLPE) {$\FLPE$}
node [draw,gray, fit = (input1) (FLPE), inner xsep = 10pt, inner ysep = 5pt,label={[text=gray]above:Alice}] (Alice) {}

node [label={$\x$},inoutput, right = of FLPE](x){\phantom{.}}
node [filter, right=of x] (HBA) {$\HB$}

node [filter, below of=HBA, node distance=\vertDist] (HEA) {$\HE$}
node [sum, right=of HEA] (sumE) {\suma}

node [sum, right=of HBA] (sumB) {\suma}
node [label={$\yB$},inoutput, right = of sumB](yB){\phantom{.}}
node [filter, right =of yB] (GB) {$\GB$}

node [filter, right =of GB, node distance=\vertDist, minimum height = 0.8cm, double] (modRedB) {$\modM$}
node [label=right:{$\nB$},inoutput, above of=sumB, node distance = 1cm](inputNB){\phantom{.}}
node [label=above:{$\rB$},inoutput, right=of modRedB](outputB){\phantom{.}}
node [draw,gray, fit = (GB) (outputB), inner xsep = 10pt, inner ysep = 5pt,label={[text=gray]above:Bob}] (Bob) {}


node [label={$\yE$},inoutput, right = of sumE](yE){\phantom{.}}
node [filter, right = of yE, node distance=\vertDist] (GE) {$\GE$}
node [filter, below of =modRedB,node distance = \vertDist, minimum height = 0.8cm, double] (modRedE) {$\modM$}
node [label=right:{$\nE$},inoutput, below of=sumE, node distance = 1cm](inputNE){\phantom{.}}
node [label=above:{$\rE$},inoutput, right=of modRedE](outputE){\phantom{.}}
node [draw,gray, fit = (GE) (outputE), inner xsep = 10pt, inner ysep = 5pt,label={[text=gray]below:Eve}] (Eve) {}
;

\draw[->, double](input1) --  (ZTx);
\draw[->, double](ZTx) -- (modRedTx);
\draw[double](modRedTx) --  (v);
\draw[->, double](v) --  (FLPE);

\draw[double](FLPE) -- (x);
\draw[->, double](x) -- (HBA);
\draw[->, double] (HBA) -- (sumB);
\draw[->, double] (inputNB) -- (sumB);
\draw[double] (sumB) -- (yB);
\draw[->, double] (yB) -- (GB);
\draw[->, double] (GB) -- (modRedB);
\draw[->, double] (modRedB) -- (outputB);

\draw[->, double] (x) -- +(0,-1) |- node {}(HEA);
\draw[->, double] (HEA) -- (sumE);
\draw[->, double] (inputNE) -- (sumE);
\draw[ double] (sumE) -- (yE);
\draw[->, double] (yE) -- (GE);
\draw[->, double] (GE) -- (modRedE);
\draw[->, double] (modRedE) -- (outputE);

\end{tikzpicture}
			\caption{\label{fig_blockDiagram}
				Block diagram of the communication setup. The left part represents
				the transmitter Alice. The top right branch corresponds to the
				legitimate receiver Bob, and the lower part to the
				eavesdropper Eve.
			}
		\end{figure*}

\noindent
		The communication scenario is depicted in the block diagram in
		Fig.~\ref{fig_blockDiagram}.
		Alice transmits data over $\NB$ parallel data streams to Bob, which allows simple processing at the legitimate receiver. 
		
		The data symbols are drawn from a \emph{QAM signal
			constellation} $\A$ with variance $\aVar$. The data symbols
		$a_j$, $j = 1,\ldots,\NB$, are combined in the information vector
		$\a \in \mathcal{A}^{\NB}$.
		It is convenient to define the signal constellation as follows
		\cite{fischer2019LatticereductionaidedIntegerforcingEqualization}.
		The signal points are a subset of points of a \emph{signal point lattice}
		$\Lambdaa = \G$, where $\G=\Z + \j\Z$ are the Gaussian integers. Using a \emph{boundary lattice}
		$\Lambdab$, which is a sublattice of  $\Lambdaa$,
		the signal constellation is defined as
		\begin{align}
			\A &= \Lambdaa \cap \mathcal{R}_\mathrm{V}( \Lambdab) \;, 
		\end{align}
		where $\mathcal{R}_\mathrm{V}(\Lambdab)$ is the \emph{Voronoi
			region} of $ \ve{\Lambda}_\mathrm{b}$ \cite{zamir2014LatticeCodingSignals}.
		Typically, an offset is included to obtain zero-mean constellations. 
		For 16-ary QAM, the boundary lattice is
		$\Lambdab = 4\G$. Including the typical offset of $t = \frac{1+\j}{2}$,
		the signal points are then of the form $a = a_\mathrm{Re} + \j\, a_\mathrm{Im}$,
		where $a_\mathrm{Re}, a_\mathrm{Im} \in \{-1.5,-0.5,0.5,1.5\}$.
		
		For each point $\x$ in the complex plane, a point within the Voronoi region
		of $\Lambdab$ can be found which is the modulo equivalent.
		This operation can be written in a linearized form as
		\cite{fischer2019LatticereductionaidedIntegerforcingEqualization}
		\begin{align}
			\modM(\x) &= \x + \ve{\lambda} \label{eq_mod_lin} \;,
		\end{align}
		where a point $\ve{\lambda} \in \Lambdab$ is added to $\x$
		such that $\x + \ve{\lambda} \in \mathcal{R}_\mathrm{V}(\Lambdab)$.

		We assume that perfect channel knowledge is available at the transmitter, i.e., Alice knows $\HB^{}$ and $\HE^{}$.
		
		Given the data vector $\a$, the transmit vector $\x$ is calculated. We consider two different structures, linear preprocessing (\LP{}) and LRA preprocessing (\LRP{}). The unified structure is displayed in Fig.~\ref{fig_blockDiagram}. The matrices $\ZLRA$ and $\FLPE$ are calculated depending on the optimization criterion.  \LP{} corresponds to a special case of \LRP{}, where $\ZLRA = \I$. The considered preprocessing techniques are presented in detail in Sec.~\ref{sec_preprocessing}.
		
		Firstly, an intermediate vector $\v$ is calculated from the information vector $\a$ according to
		\begin{align}
			\v &= \modM \left( \ZLRA^{-1} (\a - \ve{t}) \right) + \ve{t} \notag\\
			&= \ZLRA^{-1}\a + (\I - \ZLRA^{-1} )\ve{t} +\ve{\lambda} \;,
		\end{align}
		for $\ve{\lambda} \in \ve{\Lambda}_\mathrm{b}$, cf.\ (\ref{eq_mod_lin}). The usage of the offset $\ve{t} = \frac{1+\j}{2}\ve{1}$ ensures that the intermediate vector $\v$ is from the same set as $\a$. Hereby $\ve{1}$ describes the all-one vector. For a unimodular matrix $\ZLRA$, the mapping from $\a$ to $\v$ is one-to-one and thus information loss-less. Then the transmit vector is calculated using the matrix $\FLPE$,
		\begin{align}
			\x = \FLPE \v\;.
		\end{align}
		Hereby, $\FLPE$ has to be scaled suitably such that the sum power constraint $\E\{\x\x^\herm\} \leq P$ is fulfilled.
		The transmit vector is then transmitted over the channel $\HB$ to Bob and over the channel $\HE$ to Eve, respectively. Both receivers experience AWGN, described by the noise vectors $\nB$ and $\nE$.
		
		\subsection{Residual Equalization}\label{sec_eq}\noindent
			If the occurred interference is not removed by the preprocessing at the transmitter side, \emph{residual equalization} has to be performed by the receiver.
			The experienced end-to-end channels for Bob and Eve ($\bullet \in \{\mathrm{B,E}\}$ is used in the following) are given by
			\begin{align}
			\HbulT &= \Hbul \FLPE \;.
			\end{align}
			In this paper, we consider residual \emph{linear equalization (\LE{})} for both Bob and Eve. For the eavesdropper, additionally \emph{lattice decoding} using the Sphere Decoder (\SD{}) \cite{agrell2002ClosestPointSearch} is assessed. 
			
			\subsubsection{Residual Linear Equalization}
				For \LE{} according to the \emph{minimum mean-squared error (MMSE)} criterion, the equalization is performed using
				\begin{align}
				\Gbul &= \ZLRA ( \HbulT^\herm \HbulT^{} + \zeta \I)^{-1} \HbulT^\herm \label{eq_GB}
				\end{align}
				since the equalization has to be done towards $\ZLRA$ instead of $\I$. Hereby, $\zeta = \nVar/\aVar$.
				Then, the processed signal is given by \cite{fischer2019LatticereductionaidedIntegerforcingEqualization}
				\begin{align}
				\Gbul \ybul &=	\Gbul \left(\Hbul \x	+ \nbul\right)\notag\\
				&= \a + \ve{o}+\ZLRA\ve{\lambda} + \neffbul \;,
				\end{align}
				where $\ve{o} = \frac{1+\j}{2}(\ZLRA-\I)\ve{1}$ describes the added offset and $\ve{\lambda}$ is caused by the modulo reduction at the
				transmitter. To remove this periodic continuation of the original signal
				constellation, modulo reduction has to be performed by the receiver. 
				
				The performance of the receiver depends on the effective noise which comprises residual interference due to the MMSE criterion and the scaled noise. The effective noise vectors of Bob and Eve are given by
				\begin{align}
				\neffbul &= \left(\Hbul \FLPE - \ZLRA \right) \v + \Gbul \nbul\;,\label{eq_neffB}
				\end{align}
				and the effective noise variance $\neffbulVar$ is calculated according to
				\begin{align}\label{eq_varneff}
				\neffbulVar =\ &\tr{\left(\Hbul \FLPE - \ZLRA \right) \left(\Hbul \FLPE - \ZLRA \right)^\herm}  \aVar \notag\\&+ \tr{\Gbul^{}\Gbul^\herm}  \nVar\;.
				\end{align}
				The equalization is shown in the right part of Fig.~\ref{fig_blockDiagram}, where the upper part corresponds to Bob and the lower part to Eve.
			
			\subsubsection{Equalization using Lattice Decoding}
				Additionally, we also consider lattice decoding for the eavesdropper, which is implemented by the Sphere Decoder \cite{agrell2002ClosestPointSearch}. It is based on using the lattice spanned by $\HET$. Hereby, the closest lattice point to the received point $\yE$ has to be found. For this, all lattice points inside a certain sphere are investigated, until the closest point is found. We omit a more detailed description of the algorithm here, for more information see \cite{agrell2002ClosestPointSearch}.
				
				The received vector at the eavesdropper is given by
				\begin{align}
				\yE &= \HE \FLPE\v+ \nE\\
				&= \HET \v + \nE\;.
				\end{align}
				The sphere decoder delivers $\hat{\v}$, with which				
				$\hat{\a}$ is calculated according to
				\begin{align}
				\hat{\a} &= \modM(\ZLRA(\hat{\v} - \ve{t})+\ve{t})\;.
				\end{align}

		\subsection{Security Metric}\noindent
			For physical-layer security, considering the \emph{secrecy capacity} is beneficial, which corresponds to the maximum rate that can be communicated securely. The secrecy capacity is defined as \cite{leung-yan-cheong1978GaussianWireTapChannel}
			\begin{align}\label{eq_Cs}
				\Cs &= \max_\FLPE  \Rs \big( \FLPE \big) \notag \\
				&=  \max_\FLPE \Big[ \RB \big( \FLPE \big) - \RE \big( \FLPE \big) \Big]_+ \;.
			\end{align}
			Hereby, $\Rs$ is the secrecy rate, $\RB$ is the rate of the channel from Alice to Bob, $\RE$ is the rate of the channel from Alice to Eve, and $[x]_+ = \max(x,0)$.
			
			Under the assumption of Gaussian input symbols, the achievable rates of the channels from Alice to Bob and Eve, respectively, are given by 
			\begin{align}
			\Rbul \big(\FLPE \big) &= \mathrm{I}\big(\v,\ybul\big) \notag\\
			&= \log_2 \Big( \det\big( \I + \frac{\aVar}{\nVar} \Hbul^{} \FLPE^{} \FLPE^\herm \Hbul^\herm\big)\Big)\;, \label{eq_rate}
			\end{align}
			where $\mathrm{I}(.,.)$ is the mutual information. 
		
			In practical systems, the characterization by the notions
			of \emph{strong security} or the less strict \emph{weak secrecy}, cf.\ 
			\cite{bloch2011PhysicalLayerSecurity}, is infeasible as they cannot be
			derived and they are only valid for the asymptotic case of infinite block
			lengths.
			
			In \cite{klinc2011LDPCCodesGaussian}, a practical security metric, based on the error performance
			of the coding/modulation scheme is proposed. The \emph{security gap}
			$S_\mathrm{g}$ quantifies the advantage in signal-to-noise ratio (SNR) Bob
			has to have over Eve such that Bob can \emph{reliably} decode the messages,
			whereas the transmission is \emph{secure} against Eve, meaning that the
			bit error ratio (BER) of Eve is close to $0.5$. When fixing the performance
			of Bob, e.g., via transmit power control the BER is adjusted to the desired
			level, the BER of Eve can equivalently serve for characterizing the security
			level. We are aware that this is the weakest form of assessing security. In the following, the characterization of the communication scenario according to both the secrecy rate and the error ratios of Bob and Eve is considered. This results in a information-theoretic point of view as well as a practical assessment of the resulting performance.

%
%
\section{Optimal Adjustment of Preprocessing}
	\label{sec_preprocessing}
	\noindent
	Subsequently, different criteria of preprocessing are used. All preprocessing variants use the same structure of Fig.~\ref{fig_blockDiagram} ($\ZLRA = \I$ for \LP{}) and only differ in the calculation of the preprocessing matrices $\FLPE$ and $\ZLRA$. 
	
	All considered preprocessing variants are summarized in Table~\ref{tab_variants}.
	The preprocessing is optimized either using only the legitimate receiver or using knowledge about both Bob and Eve. For both variants, optimization may be done based on the error performances or the rates. For each optimization criteria, a linear and an LRA approach is presented. 
	
	For preprocessing according to the error performance of the legitimate receiver, classical linear preequalization (\LPE{}) and LRA preequalization (\LRAPE{}) are considered. Alternatively, the optimization may be done based on the rate of the channel from Alice to Bob, which we call \LPC{} or \LRPC{}.
		
	Additionally, preprocessing approaches are presented that are based on the legitimate receiver as well as the eavesdropper. Firstly, we present an approach that is based on the error performance of Bob and Eve, which is called \LPSER{} or \LRPSER{}. If the optimization is done according to the secrecy rate, the resulting preprocessing is denoted by \CsPE{} and \ZCsPE{}. 

	\begin{table}[t]
		\renewcommand{\arraystretch}{1.3}
		\caption{Considered Preprocessing Variants }
		\label{tab_variants}
		\centering
		\begin{tabular}{l|l|l|l|c}
			\hline
			Depending on &  Optimization & Structure & Name & Symbol\\%
			\hline
			\multirow{4}{*}{Only Bob} & \multirow{2}{*}{Error performance} & Linear &  \LPE{} &\inlinemark{LPEmark}{LPEcol}\\
			\cline{3-5}
			& & LRA & \LRAPE{} &\inlinemark{LRAPEmark}{LRAPEcol}\\
			\cline{2-5}
			& \multirow{2}{*}{Rate} & Linear& \LPC{} &\inlinemark{CBPEmark}{CBPEcol}\\
			\cline{3-5}
			& & LRA & \LRPC{} &\inlinemark{LRACBPEmark}{LRACBPEcol}\\
			\hline
			\multirow{4}{*}{Bob and Eve} & \multirow{2}{*}{Error performance} & Linear &  \LPSER{} &\inlinemark{SERmark}{SERcol}\\
			\cline{3-5}
			& & LRA & \LRPSER{} &\inlinemark{LRASERmark}{LRASERcol}\\
			\cline{2-5}
			& \multirow{2}{*}{Rate} & Linear & \CsPE{} &\inlinemark{CSPEmark}{CSPEcol}\\
			\cline{3-5}
			& & LRA & \ZCsPE{} &\inlinemark{LRACSPEmark}{LRACSPEcol}\\
			\hline
		\end{tabular}
		\vspace{-0.2cm}
	\end{table}
	
	\subsection{Optimization Using Only the Legitimate Receiver}\noindent
		Subsequently, optimization criteria are presented that depend only the legitimate receiver and do not require any knowledge about the eavesdropper.
		\subsubsection{Linear Preequalization (\LPE{})}
			First, we review classical linear preequalization. Here, $\ZLRA = \I$, which means  $\a = \v$ and the modulo reduction is inactive. 
			For \LPE{} according to the MMSE criterion, $\FLPE$ is given by
			\begin{align}
				\FLPE &= \frac{1}{g}\HB^\herm \left(\HB^{}\HB^\herm+\zeta\I\right)^{-1}\;,
			\end{align}
			where $\zeta = \nVar/\aVar$ and the scaling factor $\frac{1}{g}$ is used to adapt the sum transmit power to a desired value. 
					
			The legitimate receiver only has to scale suitably, residual equalization is not necessary. The equalization matrix $\GB$ is given by $\GB = g\I$. 
					
			Since the preprocessing is adapted to the channel from Alice to Bob, the eavesdropper, who experiences a different channel, has to perform residual equalization as described in Sec.~\ref{sec_eq}.

		\subsubsection{Optimization According to the Rate (\LPC{})}
			Instead of optimizing the preprocessing according to the error performance, the optimization may be performed according to the rate. Then $\FLPE$ is chosen according to
			\begin{align}
			\FLPE = \argmax_{ \Fo \in \ve{\mathcal{F}}} \RB \big(\Fo\big)\;,
			\end{align}
			where $\RB$ is calculated according to (\ref{eq_rate}).
			Hereby, a sum power constraint is considered, which means that
			\begin{align}
			\ve{\mathcal{F}} = \Big\lbrace \FLPE \mid \aVar \cdot\tr{\FLPE \FLPE^\herm} \leq P \Big\rbrace\;, \label{eq_spc}
			\end{align}
			where $P$ is the limit on the sum transmit power. This corresponds to a common constraint for wireless communications \cite{weingarten2006CapacityRegionGaussian, viswanath2003SumCapacityVector, khisti2010SecureTransmissionMultiple}.
			
		In the case of \LPC{}, the legitimate receiver does not experience an equalized channel and thus has to perform equalization, similarly to the eavesdropper. Since the goal is easy processing at the receiver, we assume that Bob performs \LE{}. For this, the received vector is multiplied by the matrix $\GB$, which is calculated according to Sec.~\ref{sec_eq}.
	
	\subsection{Optimization Using Knowledge About the Eavesdropper}\noindent
		Instead of only considering the legitimate receiver during the optimization, we also present approaches that include the eavesdropper. However, this requires knowledge about the eavesdropper which may not be available in reality.
		\subsubsection{Optimization According to the Error Performance (\LPSER{})}
			Under the assumption that the receiver performs linear equalization according to the MMSE criterion, the performance is determined by the effective noise variance, which is calculated according to (\ref{eq_varneff}) and depends on the chosen preprocessing matrix $\FLPE$. For \LPSER{}, $\FLPE$ is chosen according to
			\begin{align}
			\FLPE = \argmax_{ \Fo \in \ve{\mathcal{F}}} \neffEVar \big(\Fo\big) - \neffBVar \big(\Fo\big)\;,
			\end{align}
			where the sum power constraint according to (\ref{eq_spc}) is fulfilled.
			To recover the sent symbols, both receivers have to perform equalization, as in Sec.~\ref{sec_eq}.
			
		\subsubsection{Optimization According to the Secrecy Rate (\CsPE{})}
			The secrecy rate is given by (cf. (\ref{eq_Cs}))
			\begin{align}
				\Rs \big(\FLPE \big) = \Big[ \RB \big(\FLPE \big) - \RE \big(\FLPE \big) \Big]_+\;
			\end{align} 
			and depends on the preprocessing matrix $\FLPE$.  For \CsPE{}, preprocessing is done using
			\begin{align}
			\FLPE = \argmax_{ \Fo \in \ve{\mathcal{F}}} \Rs \big(\FLPE \big)\,
			\end{align}
			where a sum power constraint is considered (cf. (\ref{eq_spc})).
			
		As before, additional receiver-side equalization is required, as in Sec.~\ref{sec_eq}.

	\subsection{LRA Processing}\noindent
		It is advantageous to use an LRA approach, which is presented in the following.
		\subsubsection{Classical LRA Preequalization}
		For LRA preequalization (\LRAPE{}), the channel matrix is conceptually factorized into \cite{fischer2019LatticereductionaidedIntegerforcingEqualization} 
		\begin{align}						     \label{eq_Hfactor}
		\HB &= \ZLRA\ve{W} \;,
		\end{align}
		where $\ve{W}$ corresponds to non-integer interference and the full-rank Gaussian integer matrix $\ZLRA$ describes integer interference. $\FLPE$ is then given by
		\begin{align}
		\FLPE = \frac{1}{g} \HB^\herm \left( \HB^{} \HB^\herm + \zeta \I \right)^{-1} \ZLRA \;.\label{eq_FLPE}
		\end{align}
		
		For best performance, the factorization is not performed based on (\ref{eq_Hfactor}), but using the augmented matrix $\ve{\mathcal{H}}_\mathrm{BA}$ \cite{fischer2019LatticereductionaidedIntegerforcingEqualization}
		
		\begin{align}
		\HAug &= \begin{bmatrix} \HB & \sqrt{\zeta}\I \end{bmatrix}
		\end{align}
		and $\ZLRA$ is obtained according to
		\begin{align}
		\ZLRA_\mathrm{opt} = \argmin_{\ZLRA \in \G^{\NB \times \NB}} || \begin{bmatrix}
		\HB & \sqrt{\zeta}\I
		\end{bmatrix}^+\, \ve{Z}^\herm ||^{}_\mathrm{F} \;,
		\end{align}
		where $(.)^+$ denotes the pseudo-inverse. For details on the factorization, see \cite{stern2017OptimalFactorizationLatticeReductionAided}.

		Similarly to \LPE{}, the legitimate receiver only has to scale the received signal suitably and perform modulo reduction. The eavesdropper has to perform equalization to remove the interference occurred due to the preprocessing and the transmission over the channel, see Sec.~\ref{sec_eq}.

		\subsubsection{LRA Processing for Fixed $\FLPE$}
		
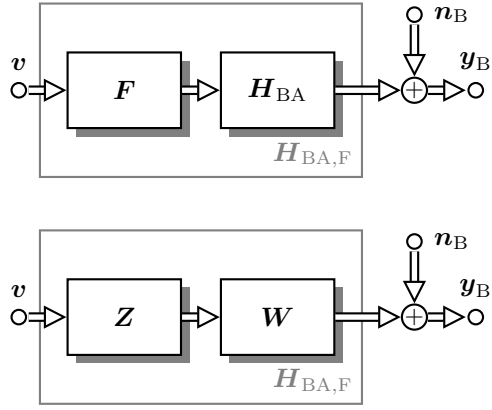
\begin{figure}[t]
				\centering

\begin{tikzpicture}[auto, thick, node distance=\blockDist, >={Triangle[open,length=3mm,width=3mm]}, double distance = 2pt]
\draw
node at (0,0)[label={$\v$},inoutput](v){\phantom{.}}

node [filter, right=of v] (FLPE) {$\FLPE$}

node [filter, right=of FLPE] (HBA) {$\HB$}
node [draw,gray, fit = (FLPE) (HBA), inner xsep = 10pt, inner ysep = 18pt] (HBF) {}
node[above left, text=gray] at (HBF.south east) {$\HBT$}
node [sum, right=of HBF] (sumB) {\suma}
node [label={$\yB$},inoutput, right = of sumB](yB){\phantom{.}}

node [label=right:{$\nB$},inoutput, above of=sumB, node distance = 1cm](inputNB){\phantom{.}}

node at (0,-3)[label={$\v$},inoutput](v2){\phantom{.}}
node [filter, right=of v2] (Z) {$\ZLRA$}

node [filter, right=of Z] (W) {$\W$}
node [draw, fit = (Z) (W), inner xsep = 10pt, inner ysep = 18pt, behind path, gray] (HBF2) {}
node[above left, text=gray] at (HBF2.south east) {$\HBT$}

node [sum, right=of HBF2] (sumB2) {\suma}
node [label={$\yB$},inoutput, right = of sumB2](yB2){\phantom{.}}

node [label=right:{$\nB$},inoutput, above of=sumB2, node distance = 1cm](inputNB2){\phantom{.}}
;


\draw[->, double](v) --  (FLPE);

\draw[->, double](FLPE) -- (HBA);
\draw[->, double] (HBA) -- (sumB);
\draw[->, double] (inputNB) -- (sumB);
\draw[->, double] (sumB) -- (yB);

\draw[->, double](v2) --  (Z);

\draw[->, double](Z) -- (W);
\draw[->, double] (W) -- (sumB2);
\draw[->, double] (inputNB2) -- (sumB2);
\draw[->, double] (sumB2) -- (yB2);

\end{tikzpicture}
				\caption{Lattice reduction is done based on the cascade of preprocessing and transmission channel, which results in the integer matrix $\ZLRA$ and the reduced basis $\W$. \label{fig_block_LRA}}
			\end{figure}

			We also present an LRA approach for the case that the preprocessing matrix $\FLPE$ is fixed. This approach may be used to allow LRA processing if $\FLPE$ has been calculated according to any of criteria that have been described before. In this case, the lattice reduction is based on the cascade of transmission channel and preprocessing $\HBT = \HB\FLPE$, which is shown in the top part of Fig.~\ref{fig_block_LRA}.
		Using lattice reduction, $\HBT$ is conceptually decomposed into
		\begin{align}
		\HBT = \W \ZLRA\;,\label{eq_Hfactor2}
		\end{align}
		where $\ZLRA$ is a full-rank Gaussian integer matrix and $\W$ corresponds to the non-integer interference. This factorization is done as for LRA equalization based on the augmented matrix $\HAug$ and is the dual to the preequalization (cf. (\ref{eq_Hfactor})). Then, the non-integer interference is equalized at the receiver using $\GB = \W^+$, while the influence of $\ZLRA$ is canceled out due to the multiplication of $\ZLRA^{-1}$ at the transmitter. This approach is used for \LRPC{}, \LRPSER{}, and \ZCsPE{}.
			
		\subsubsection{Channel Knowledge} \label{sec_channel_knowledge} 
	
	As shown in \cite{schulz2023LatticeReductionAidedPreequalizationPhysicalLayer}, the eavesdropper is not able to gain the knowledge about the matrix $\ZLRA$ in case of \LRAPE{}. As shown in Sec.~\ref{sec_eq}, an estimate of $\ZLRA$ is required for the equalization. Without this knowledge, the original data symbols cannot be recovered.
	
	To be able to calculate the preprocessing, knowledge about $\HB$ has to be available at the transmitter. For this, training sequences are sent in a startup phase before the actual transmission. We assume that the channel is reciprocal, hence Alice is able to measure the channel $\ve{H}_\mathrm{AB}$ using training sequences sent by Bob and uses this measurement as an estimate of $\HB$. This enables the eavesdropper to estimate the channel $\ve{H}_\mathrm{BE}$. However this knowledge cannot be exploited by the eavesdropper.
	
	If a training sequence is sent by Alice, the eavesdropper is able to measure $\HE$.
	For the transmission, Alice activates the preprocessing. Using a blind estimator, Eve may try to estimate the end-to-end channel. However, the eavesdropper is only able estimate $\HET = \HE \FLPE$, which allows her to recover the intermediate symbols $\v$. If $\ZLRA = \I$, this means that Eve can recover the original data symbols, since $\a = \v$. 
	
	Since $\v \neq \a$ for $\ZLRA \neq \I$, the eavesdropper can only calculate $\v$, which corresponds to linear combinations of $\a$. To calculate $\a$ from $\v$, the matrix $\ZLRA$ has to be available. The intermediate vector $\v$ and the data symbols $\a$ consist of the same symbols, which makes it impossible for the eavesdropper to gain an estimate of the mapping from $\a$ to $\v$ and thus, an estimate of $\ZLRA$.
	
	If $\HE$ and $\HET$ are known to Eve and $\NE \geq \NA$, Eve may calculate
	\begin{align}
	\HE^+\HET^{} &= \HE^+\HE^{} \FLPE =\FLPE\;,
	\end{align} 
	where $\HE^+ = (\HE^\herm \HE)^{-1}\HE^\herm$ is the pseudo-inverse of $\HE$.
	
	If \LRAPE{} is used, the pseudo-inverse of $\FLPE$ is
	\begin{align}
	\FLPE^+ &= \left(\HB^\herm\left(\HB\HB^\herm+\zeta\I\right)^{-1}\ZLRA\right)^+ \notag \\
	&= \ZLRA^{-1} \left(\I+\zeta\left(\HB\HB^\herm\right)^{-1}\right)\HB \notag\\
	&\approx \ZLRA^{-1}\HB = \ve{W}\;,
	\end{align}
	which corresponds to the reduced channel $\ve{W}$, cf. (\ref{eq_Hfactor}). 
	There are an infinite number of original channels $\HB$ and corresponding integer matrices $\ZLRA$. Hence, $\ZLRA$ cannot be calculated from the estimates that Eve is able to obtain.
	
	For the \LRPC{}, \LRPSER{}, and \LRPSER{}, also no knowledge of $\ZLRA$ can be gained given $\FLPE$. The factorization is based on $\HBT$, which means that knowledge about $\HB$ is required to perform the factorization, since $\HB\FLPE = \W\ZLRA$, cf. (\ref{eq_Hfactor2}). With only knowledge about $\FLPE$ available, there are an infinite number of possibilities to fulfill this.
	
	 Consequently, if $\ZLRA \neq \I$ Eve is only able to recover the intermediate vector $\v$ and not the original data symbols $\a$. Thus, the matrix $\ZLRA$ can be seen as a secret key that is calculated from $\HB$ and not known to the eavesdropper.

%
%

\section{Numerical Results}
	\label{sec_numericalResults}
	\noindent
	We now present results from numerical simulations. The simulation parameters are presented in Table~\ref{tab_params}.

	\subsection{Simulation Parameters}
	\noindent
	Alice wants to communicate 16-ary QAM symbols to Bob by performing
	preprocessing as described in Sec.~\ref{sec_preprocessing}. All antennas used in this scenario have a gain of $15\unit{dBi}$ and a half power bandwidth of $33^{\circ}$. We assume that parabolic antennas are used to calculate the angle-dependent antenna gains according to \cite[Eq.~(4.84)]{collin1985AntennasRadiowavePropagation}.
	
	The transmitter Alice is placed under the ceiling in the middle of the hall in the position $\begin{bmatrix} 0 & 0 & 3\unit{m} \end{bmatrix}$ with a fixed alignment of the $\NA = 3$ antennas towards Bob. The transmitter antennas are arranged in a right triangle with a distance of 3.06\unit{cm} and 3.07\unit{cm} between the antennas. This accounts for non-perfect positioning of the antennas which occurs due to the small size. 
	
	The legitimate receiver Bob is directly below the transmitter in position $\begin{bmatrix} 0 & 0 & 0 \end{bmatrix}$. The antenna spacing for Bob and Eve is 3\unit{cm}. For both receivers multiple rotations of the antenna array are considered. In reality a perfect alignment is hard to achieve and different alignments occur. While the height of the antennas stays constant, the arrays are rotated in the horizontal plane. The considered rotations of the antenna arrays are chosen randomly from a uniform distribution in the interval $[0,\pi/2]$. The preprocessing is calculated depending on the rotation of the arrays. 
	 
		\begin{table}[t]
		\renewcommand{\arraystretch}{1.3}
		\caption{Parameters for Numerical Simulations. }
		\label{tab_params}
		\centering
		\begin{tabular}{lll}
			\hline
			Parameter &  Variable & Value\\
			\hline
			Carrier frequency & $f_\mathrm{c}$&300\unit{GHz}\\
			Bandwidth & $B$ & 3\unit{GHz}\\
			Temperature & $T$ & 290\unit{K}\\
			Noise figure & $F$ & 9\unit{dB}\\
			Signal constellation & $\mathcal{A}$& 16-QAM\\
			Antenna gain & $G_\mathrm{A},G_\mathrm{B},G_\mathrm{E}$& 15\unit{dBi}\\
			Number of transmit antennas & $\NA$& 3\\
			Number of receive antennas & $\NB,\NE$& 2\\
			Position of Alice & & $[0,0,3\unit{m}]$\\
			Position of Bob & & $[0,0,0]$\\
			Position of $\text{Eve}_1$ &  & $[1\unit{m}, 1\unit{m}, 0]$\\
			Position of $\text{Eve}_2$ &  & $[2\unit{m}, 2\unit{m}, 0]$\\
			\hline
		\end{tabular}
		\vspace{-0.2cm}
	\end{table}
	For the evaluation of the results, the worst case assumption is considered that Eve always chooses the best possibility to achieve the best results. Hence, of the randomly chosen rotations, the best rotation of Eve is considered for each rotation of Bob. Then the mean of all results is taken over the different rotations of Bob to achieve more general results. 
	
	The channel matrices are calculated as described in Sec.~\ref{subsec_channelModel}. We assume that perfect channel knowledge is available at all communication partners. As described in Sec.~\ref{sec_channel_knowledge}, this is a worst case assumption if $\ZLRA \neq \I$, since the eavesdropper may not be able to obtain all necessary knowledge in practice. 
	
	All optimizations are done using a numerical optimization algorithm based on the interior-point method. Due to the non-convexity, the global optimum cannot be guaranteed. However, the optimization is done using multiple different initial points such that the found solution is the global optimum with high probability.
	\subsection{Exemplary Positions}\noindent
	Firstly, we consider an exemplary position of an eavesdropper, who is placed at the same height as Bob with a small distance in position $[ 1\unit{m}, 1\unit{m}, 0 ]$. For the exemplary positions 100 random rotations are considered for the antenna arrays of Eve and 800 different rotations for Bob.
		\subsubsection{Optimization Using Only the Legitimate Receiver}
\begin{figure}[t]
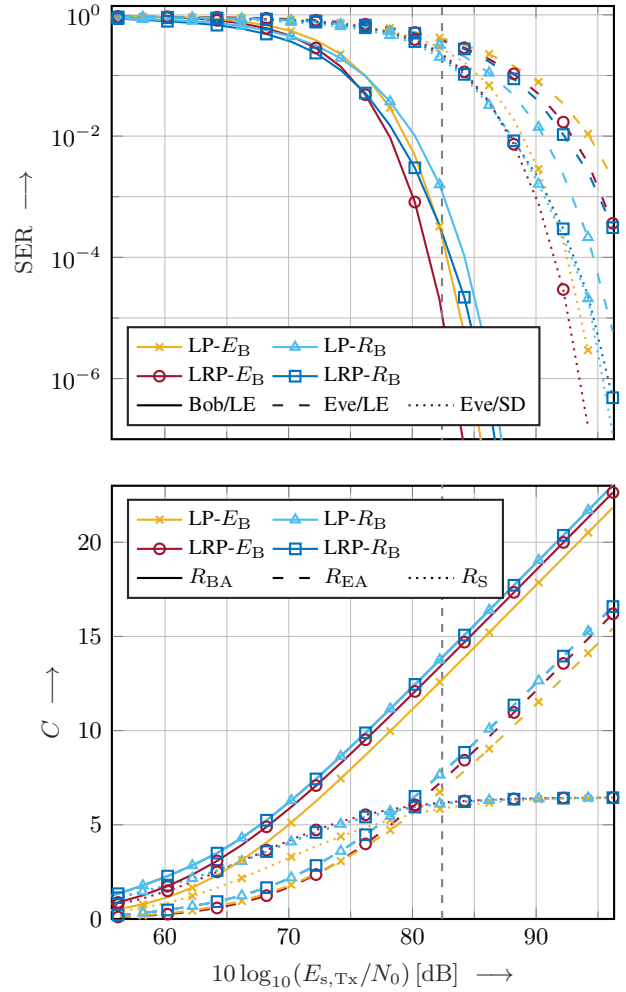

				\flushright
				\subfloat
					{	\input{SNR_SER_LPE_CB_LRA.tikz}\hspace{0.8cm}}

				\subfloat
					{\input{SNR_CGauss_LPE_CB_LRA.tikz}\hspace{0.8cm}}
				\caption{Comparison of \LPE{}, \LRAPE{}, \LPC{}, and \LRPC{} over the SNR for Bob 	and $\text{Eve}_1$. Top: SER, bottom: rates and secrecy rate. \label{fig_LPE_CB}}
			\end{figure}

			In the following, we present the results for the optimization variants that only consider Bob. The top part of Fig.~\ref{fig_LPE_CB} shows the resulting SER of \LPE{}, \LRAPE{}, \LPC{}, and \LRPC{} over the SNR.
			There is a significant gain in the performance of Bob compared to that of Eve. As expected, employing the \SD{} results in a lower SER than using \LE{}.
			For both optimization criteria, a considerable improvement is achieved by the LRA variant compared to linear preprocessing. In case of preprocessing according to Bob's rate, only the legitimate receiver and the eavesdropper using \LE{} achieve a gain of \LRPC{} compared to \LPC{}, while using \SD{}, the same performance is achieved for both variants. For high SNR, the lowest SER of Bob is achieved by \LRAPE{}, while for low SNR \LRPC{} achieves a small gain. 
			
			The lower part of Fig.~\ref{fig_LPE_CB} shows the corresponding rates of the channels from Alice to Bob $\RB$, from Alice to Eve $\RE$, and the secrecy rates $\Rs$. Since for \LPC{} and \LRPC{} the same matrix $\FLPE$ is used, the same rates are achieved for both variants. A small gain of $\RB$ and $\RE$ is visible compared to \LPE{} and \LRAPE{}. The secrecy rates achieved by \LPC{}, \LRPC{}, and \LRAPE{} show a small gain compared to \LPE{}. When comparing \LRAPE{} to \LPE{}, it can be seen that an improvement is achieved for $\RB$, $\RE$, and $\Rs$ for the LRA variant. 
			
			All these variants do not require any knowledge about the eavesdropper and only depend on the legitimate receiver. It can be seen that while a gain is achieved for Bob, the eavesdropper also experiences low SER for high SNR.
		
		\subsubsection{Optimization Using Knowledge About the Eavesdropper}
\begin{figure}[t]
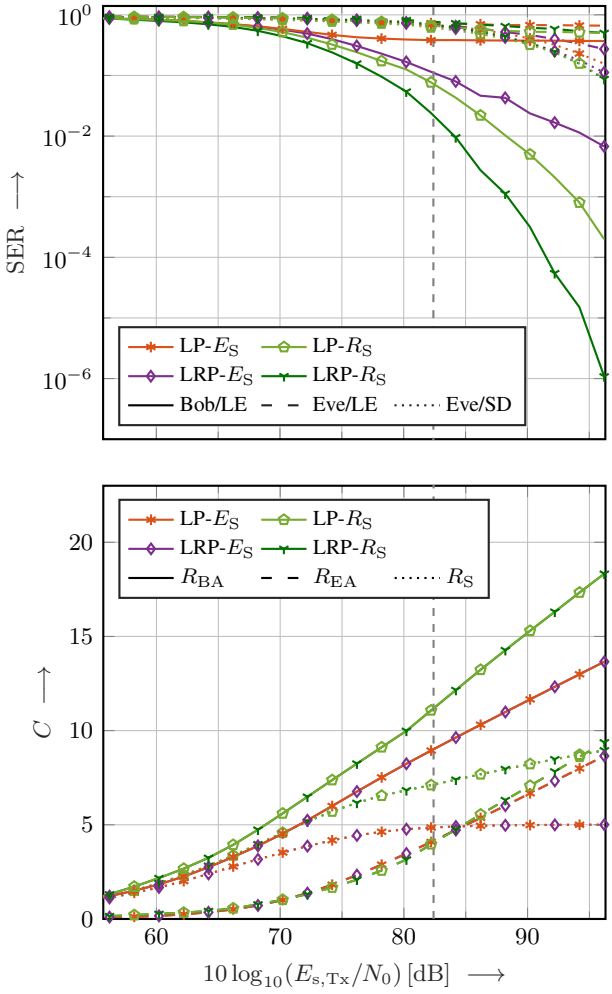

				\flushright
				\subfloat
					{	\input{SNR_SER_SER_Cs_LRA.tikz}\hspace{0.8cm}}
				\hfil
				\subfloat
					{\input{SNR_CGauss_SER_Cs_LRA.tikz}\hspace{0.8cm}}
				\caption{Comparison of \LPSER{}, \LRPSER{}, \CsPE{}, and \ZCsPE{} over the SNR for Bob and $\text{Eve}_1$. Top: SER, bottom: rates and secrecy rate. \label{fig_SER_Cs}}
			\end{figure}

			Now the approaches are presented that include the security directly into the design of the preprocessing. For the optimization, knowledge about the eavesdropper is required. 
			In the top part of Fig.~\ref{fig_SER_Cs} the SER of \LPSER{}, \LRPSER{}, \CsPE{}, and \ZCsPE{} of Bob and $\text{Eve}_1$ are displayed. 
			For the eavesdropper, only a small gain is achieved by using \SD{} in comparison to \LE{}, which shows almost no improvement even for high SNR. For the eavesdropper, the differences in the error rates of the different preprocessing variants are only small, whereas for the legitimate receiver larger differences occur. 
			For both optimization criteria, the LRA variant results in a significant improvement compared to the linear preprocessing. The best performance is achieved by \ZCsPE{}. For \LPSER{}, Bob's performance shows almost no improvement over the SNR and the SER is very high. Similarly, if the eavesdropper uses \LE{}, the resulting SER is very high and no improvement for high SNR is visible. Detection using the \SD{}, an improvement is possible, and the resulting performance even surpasses that of the legitimate receiver.
			
			In the lower part of Fig.~\ref{fig_SER_Cs}, the resulting rates and secrecy rates are depicted. For both optimization criteria, the same matrices $\FLPE$ are used for the linear and the LRA preprocessing, thus, the resulting rates are the same. While $\RE$ is very similar for both optimization criteria, $\RB$ shows a significant gain for \CsPE{} and \ZCsPE{} compared to \LPSER{} and \LRPSER{}. Consequently, the resulting secrecy rates also show an improvement.
			
		\subsection{Security vs. Reliability}
\begin{figure}
				\flushright

%
\begin{tikzpicture}[trim axis right, trim axis left, baseline]

\begin{axis}[%
ymode=log,
SERCsStyle,
legend columns=4,
mark size=2.5pt
]

\addplot[only marks, mark=., mark options={},  draw=white, mark size=0.1pt,forget plot] 
plot coordinates{(13.9318904494294,	0.00105216289473684)} node[pos=0](B2){};

\addplot[only marks, mark=., mark options={},  draw=white, mark size=0.1pt,forget plot] 
plot coordinates{(6.18238147246704,	1)} node[pos=0](B){};

\addplot[only marks, mark=LRAPEmark, mark options={},  draw=LRAPEcol, forget plot] 
plot coordinates{(6.18238147246704,	1.045125e-05)} node[pos=0](A){};

\addplot[only marks, mark=LPEmark, mark options={},  draw=LPEcol, forget plot] 
table[row sep=crcr]{%
	x	y\\
	5.87610876759101	0.00020309375\\
};

\addplot[only marks, mark=LRAPEmark, mark options={},  draw=LRAPEcol, forget plot] 
plot coordinates{(13.5544494241904,	1.034975e-05)} node[pos=0](A2){};

\addplot[only marks, mark=LPEmark, mark options={},  draw=LPEcol, forget plot] 
table[row sep=crcr]{%
	x	y\\
	12.7577124661808	0.00020294175\\
};

\addplot[only marks, mark=LRACBPEmark, mark options={},  draw=LRACBPEcol, forget plot] 
table[row sep=crcr]{%
	x	y\\
	6.12974127143469	0.00023948625\\
};

\addplot[only marks, mark=CBPEmark, mark options={},  draw=CBPEcol, forget plot] 
table[row sep=crcr]{%
	x	y\\
	6.12974130759728	0.00108907925\\
};

\addplot[only marks, mark=LRACBPEmark, mark options={},  draw=LRACBPEcol, forget plot] 
table[row sep=crcr]{%
	x	y\\
	13.932312704164	0.000245038\\
};

\addplot[only marks, mark=CBPEmark, mark options={},  draw=CBPEcol, forget plot] 
table[row sep=crcr]{%
	x	y\\
	13.9323126898369	0.001053472\\
};

\addplot[only marks, mark=LRASERmark, mark options={},  draw=LRASERcol, forget plot] 
table[row sep=crcr]{%
	x	y\\
	4.86264781421925	0.110358849875\\
};

\addplot[only marks, mark=SERmark, mark options={},  draw=SERcol, forget plot] 
table[row sep=crcr]{%
	x	y\\
	4.86264723260057	0.388364987125\\
};

\addplot[only marks, mark=LRASERmark, mark options={},  draw=LRASERcol, forget plot] 
table[row sep=crcr]{%
	x	y\\
	13.4455577520089	0.000117035375\\
};

\addplot[only marks, mark=SERmark, mark options={},  draw=SERcol, forget plot] 
table[row sep=crcr]{%
	x	y\\
	13.4455577579538	0.00011272025\\
};

\addplot[only marks, mark=LRACSPEmark, mark options={},  draw=LRACSPEcol, forget plot] 
table[row sep=crcr]{%
	x	y\\
	7.14711913253948	0.02144351\\
};

\addplot[only marks, mark=CSPEmark, mark options={},  draw=CSPEcol, forget plot] 
table[row sep=crcr]{%
	x	y\\
	7.14711947423488	0.07146010925\\
};

\addplot[only marks, mark=LRACSPEmark, mark options={},  draw=LRACSPEcol, forget plot] 
table[row sep=crcr]{%
	x	y\\
	13.931890474698	0.000186608815789474\\
};

\addplot[only marks, mark=CSPEmark, mark options={},  draw=CSPEcol, forget plot] 
table[row sep=crcr]{%
	x	y\\
	13.9318904494294	0.00105216289473684\\
};

\node (C)[very thick,draw=gray!50!white,ellipse,rotate fit=-0,, inner xsep = 18pt, inner ysep = -10pt, fit=(A) (B), label={[text=gray!70!white]above right:$\text{Eve}_1$}, ] {};
\node (C)[very thick,draw=gray!50!white,ellipse,rotate fit=-0,, inner xsep = 8pt, inner ysep = 0pt, fit=(A2) (B2), label={[text=gray!70!white]above left:$\text{Eve}_2$}] {};

\addlegendimage{color=LPEcol, mark=LPEmark, only marks}
\addlegendentry{ \LPE{}}

\addlegendimage{color=CBPEcol, mark=CBPEmark, only marks}
\addlegendentry{ \LPC{}}

\addlegendimage{color=SERcol, mark=SERmark, only marks}
\addlegendentry{ \LPSER{}}

\addlegendimage{color=CSPEcol, mark=CSPEmark, only marks}
\addlegendentry{ \CsPE{}}

\addlegendimage{color=LRAPEcol, mark=LRAPEmark, only marks}
\addlegendentry{ \LRAPE{}}

\addlegendimage{color=LRACBPEcol, mark=LRACBPEmark, only marks}
\addlegendentry{ \LRPC{}}

\addlegendimage{color=LRASERcol, mark=LRASERmark, only marks}
\addlegendentry{ \LRPSER{}}

\addlegendimage{color=LRACSPEcol, mark=LRACSPEmark, only marks}
\addlegendentry{ \ZCsPE{}}

s\end{axis}
\end{tikzpicture}%
	\hspace{0.8cm}
				\caption{Secrecy rates and Bob's SER of different preprocessing variants for $\text{Eve}_1$ and $\text{Eve}_2$. \label{fig_Sec_Rel}}
			\end{figure}
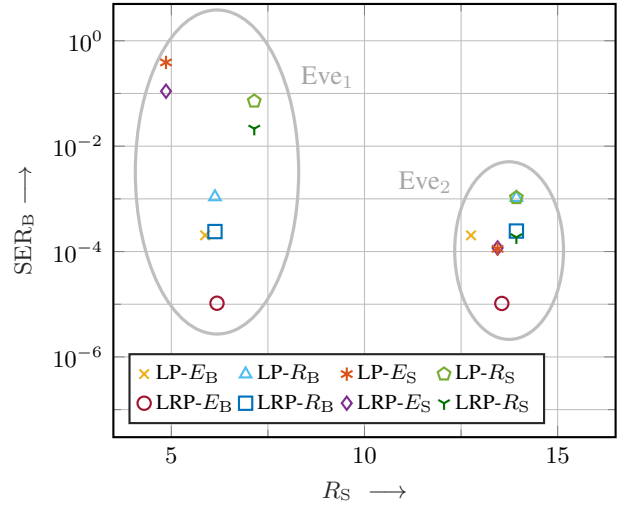

\noindent
			The goal of the communication is to achieve a secure and reliable transmission. To evaluate both aims, we assess the achieved secrecy rate together with Bob's SER. The SNR is fixed to a value of $10\log_{10}(\EsT/\No) = 82.5 \unit{dB}$, which is depicted by gray dashed lines in Fig.~\ref{fig_LPE_CB} and Fig.~\ref{fig_SER_Cs}. Here, two different positions of the eavesdropper are considered, $\text{Eve}_1$ is in position $[1\unit{m}, 1\unit{m}, 0 ]$ and $\text{Eve}_2$ is placed at $[2\unit{m}, 2\unit{m}, 0]$. The results for the different preprocessing variants are shown in Fig.~\ref{fig_Sec_Rel}.
		
			Since \LPE{}, \LRAPE{}, \LPC{}, and \LRPC{} are calculated independently of the eavesdropper, the resulting error ratios of Bob are the same for $\text{Eve}_1$ and $\text{Eve}_2$. For all optimization criteria, the SER of Bob is improved by using the LRA variant. Only for $\text{Eve}_2$, the same results are achieved for \LPSER{} and \LRPSER{}. While the secrecy rate is the same for the other optimization criteria, a gain can be achieved by using \LRAPE{} compared to \LPE{}, which is increased for $\text{Eve}_2$.
			
			For both optimization criteria that include the eavesdropper, Bob's SER shows a considerable improvement for $\text{Eve}_2$, who is placed at a larger distance. For $\text{Eve}_2$ both the SER of Bob and the secrecy rate show the same values for optimization according to Bob's rate and the secrecy rate, while there are large differences in the results for $\text{Eve}_1$.

	\subsection{Security Maps}\noindent
		To evaluate the corresponding performance of the eavesdropper, we also employ so called \emph{security maps} \cite{utkovski2019LearningRadioMaps}. They show the performance for multiple positions of an eavesdropper and thus provide a more general overview. Bob and Alice are in the positions as specified in Table~\ref{tab_params}. For the optimization of \ZCsPE{} and \LRPSER{}, the position of $\text{Eve}_1$ is used. Subsequently, the resulting performance is evaluated for multiple positions that are distributed over the area, where the eavesdropper uses \SD{}. The SER or secrecy rate corresponding to eavesdroppers in different positions is indicated by the color of the respective point. For Bob, 500 different rotations are considered, while for the eavesdropper the best of 20 rotations is chosen. In each scenario the transmit power is chosen such that the legitimate receiver experiences an SER of $\approx 5 \cdot 10^{-4}$. Consequently, the transmit power varies for each variant. While \LRAPE{} and \LRPC{} require the lowest transmit power to achieve the target SER, a significant increase in transmit power is necessary in case of \ZCsPE{} and \LRPSER{}.

		\begin{figure}[tb]
			\centering
			\begin{tikzpicture}[node distance = 0.1cm]
			\node[inner sep=0] (picLRAPE) at (0,0) {\includegraphics[height=\mapHeightS,trim= {65 0 162 15}, clip]
				{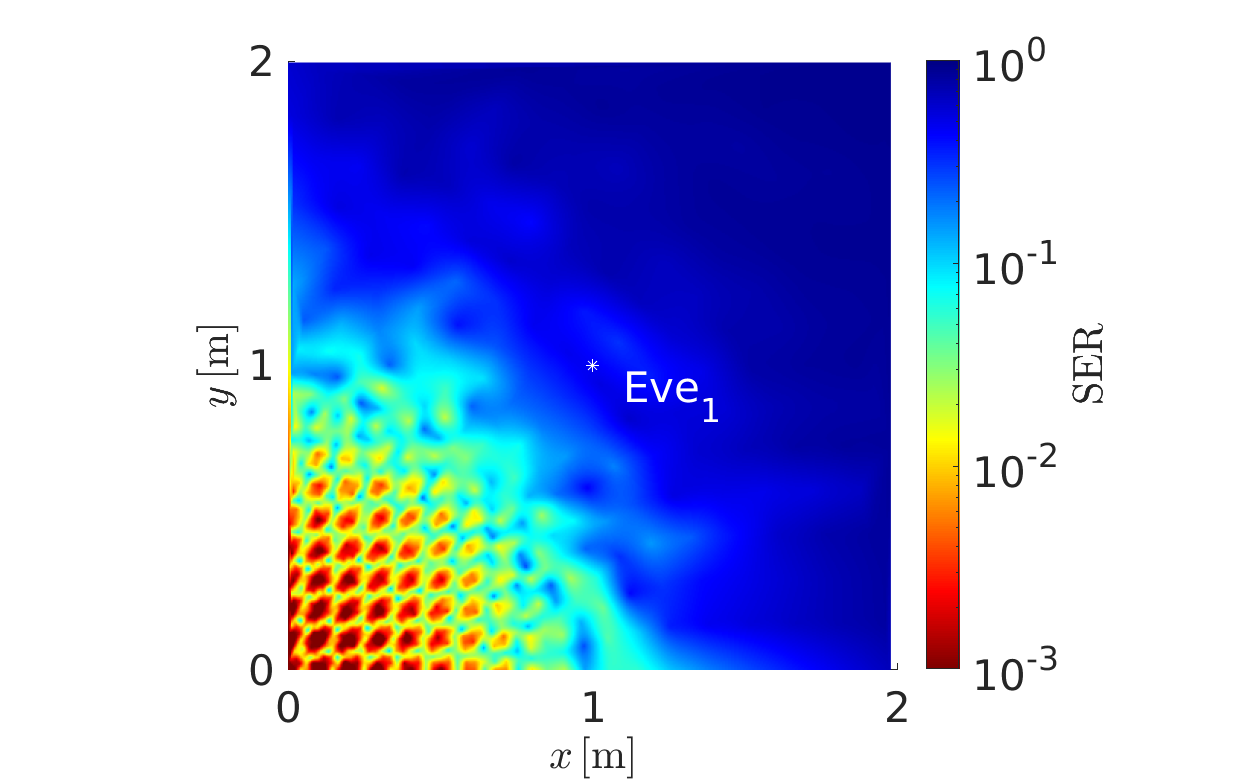}};
			\node[label={[text=white]below left:\LRAPE{} }] at (picLRAPE.north east) {};
			
			\node[right = of picLRAPE,inner sep=0] (picSER){\includegraphics[height=\mapHeightS,trim= {65 0 162 15}, clip]
				{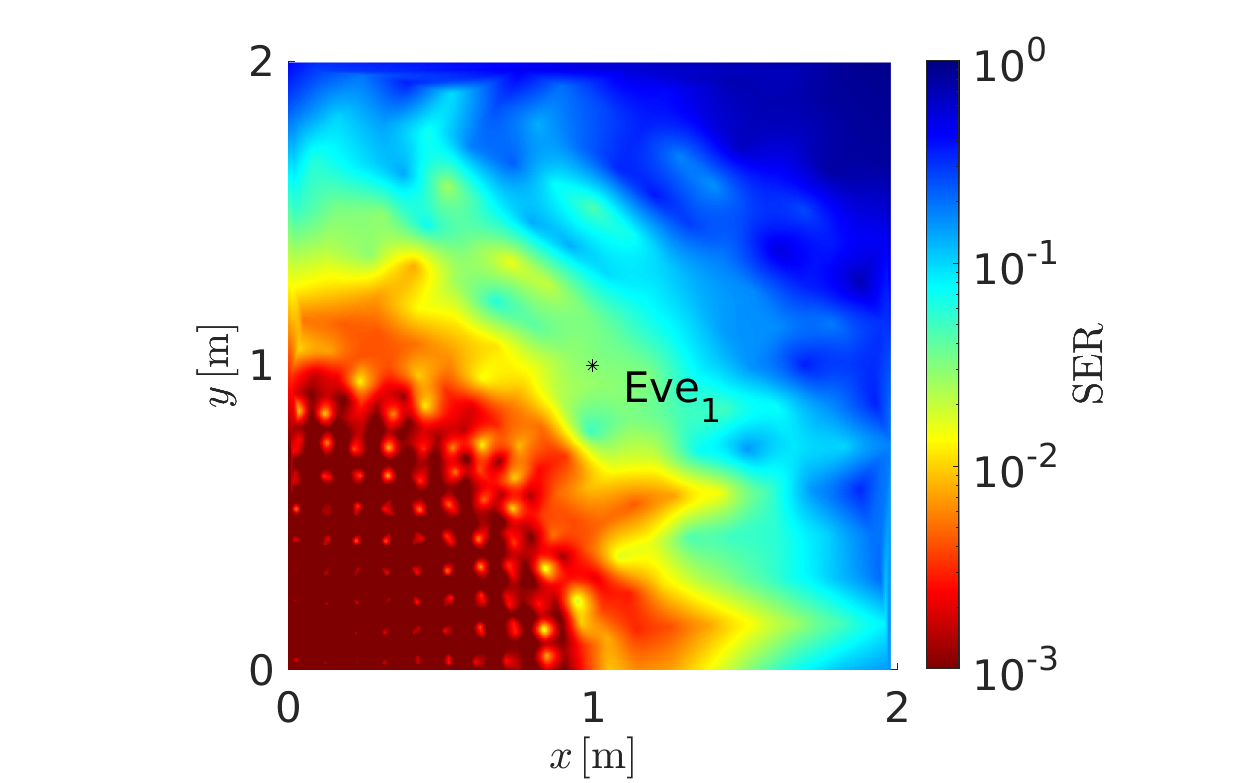}};
			\node[label={[text=white]below left: \LRPSER{} }] at (picSER.north east) {};
			
			\node[below = of picLRAPE,inner sep=0] (picCB){\includegraphics[height=\mapHeightS,trim= {65 0 162 15}, clip]
				{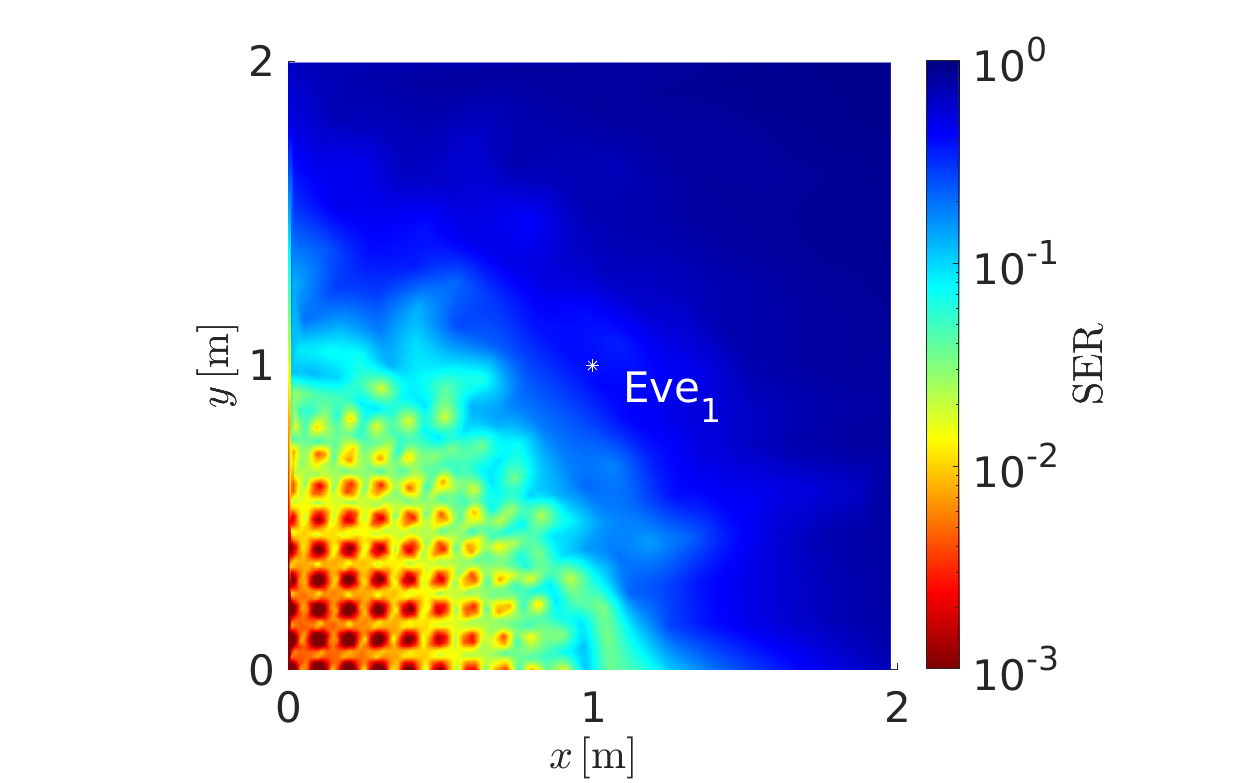}};
			\node[label={[text=white]below left:  \LRPC{} }] at (picCB.north east) {};
			
			\node[right = of picCB,inner sep=0, node distance = 0.5cm] (picCs){\includegraphics[height=\mapHeightS,trim= {65 0 60 15}, clip]
				{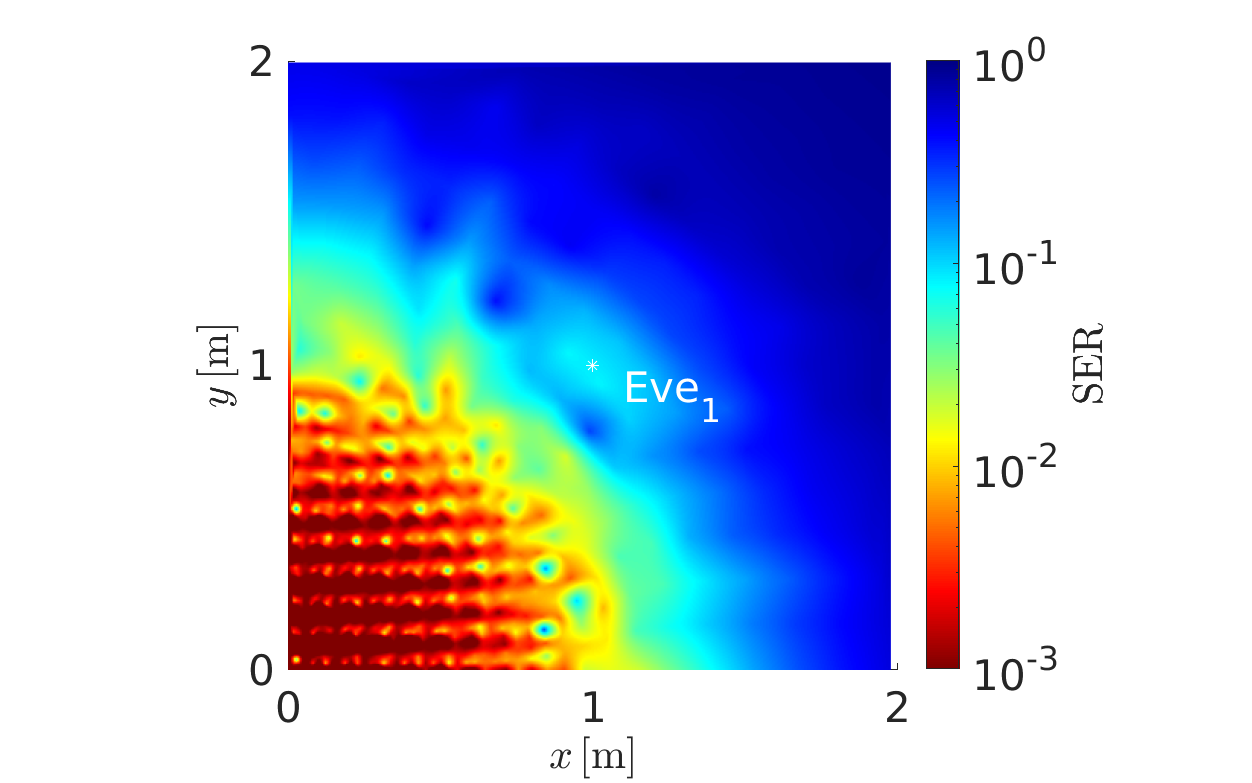}};
			
			\node[right = of picCB,inner sep=0] (picCsinv){\includegraphics[height=\mapHeightS,trim= {65 0 162 15}, clip]
				{135_h3_Atri_m3_G15_B1_2_R500_G15_E1084_2_R20_G15_B3_nF1_19dB_sims1_n20000_CsOptLRAGaussGaussIMin10Matlab_ser_minERot_SD}};
			\node[label={[text=white]below left: \ZCsPE{} }] at (picCsinv.north east) {};
			\end{tikzpicture}
			
			\caption{Comparison of the SER different preprocessing variants. \label{fig_SERMaps}}
		\end{figure}
	
		Fig.~\ref{fig_SERMaps} shows the resulting SER of the eavesdropper for \LRAPE{}, \LRPSER{}, \LRPC{}, and \ZCsPE{}. For large distances of the eavesdropper, a very high SER is achieved for all variants. For small distances, there is a pattern visible that is similar for all variants. Consequently, a small change in the position of the eavesdropper may result in a large increase or decrease of the SER. Interestingly, both variants that do not include the eavesdropper in the optimization achieve the fewest points that result in a low SER of the eavesdropper. While \ZCsPE{} and \LRPSER{} also consider the eavesdropper in the optimization, the legitimate receiver also suffers in performance. Since in this scenario Bob's target SER is fixed, a higher transmit power is required to fulfill this. However, the eavesdropper also benefits from the increase in transmit power. Thus, more positions show a green or red color and thus experience a lower SER compared to \LRAPE{} and \LRPC{}.
		\begin{figure}[tb]
			\centering
			\begin{tikzpicture}[node distance = 0.1cm]
			\node[inner sep=0] (picLRAPE) at (0,0) {\includegraphics[height=\mapHeightS,trim= {65 0 162 15}, clip]
				{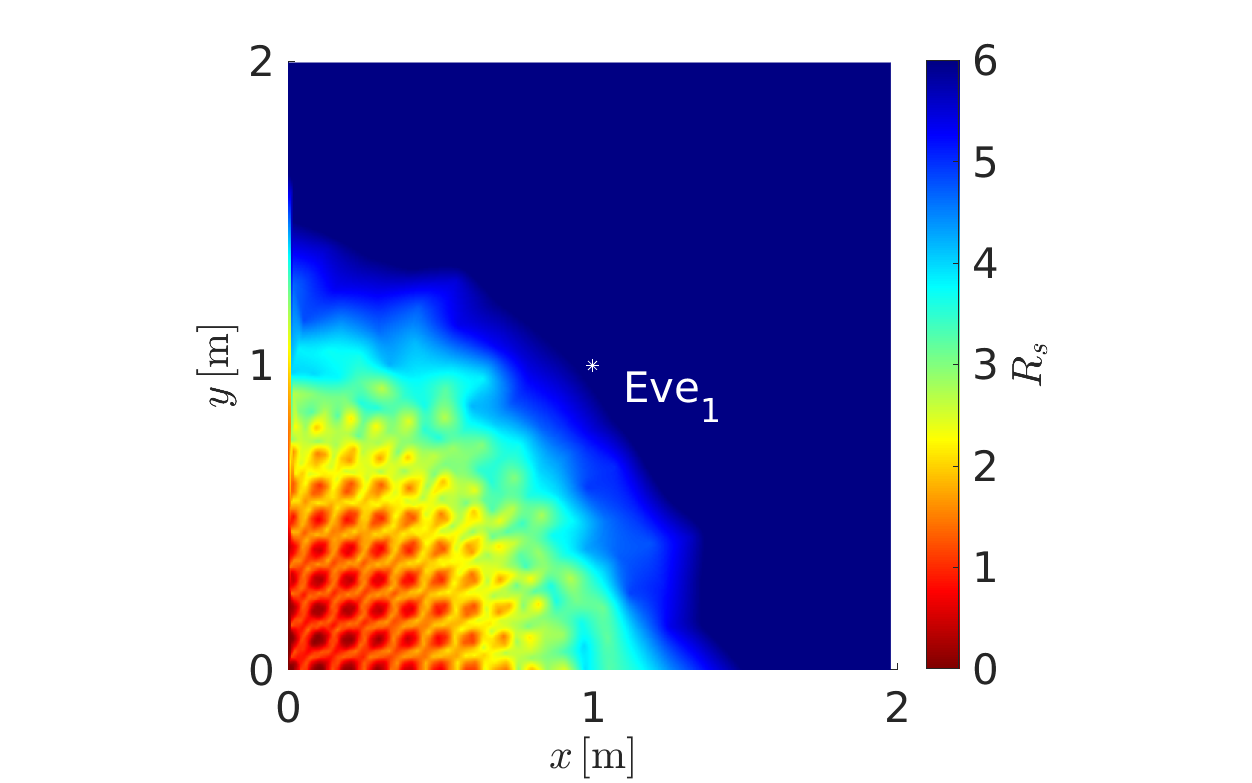}};
			\node[label={[text=white]below left:\LRAPE{} }] at (picLRAPE.north east) {};
			
			\node[right = of picLRAPE,inner sep=0] (picSER){\includegraphics[height=\mapHeightS,trim= {65 0 162 15}, clip]
				{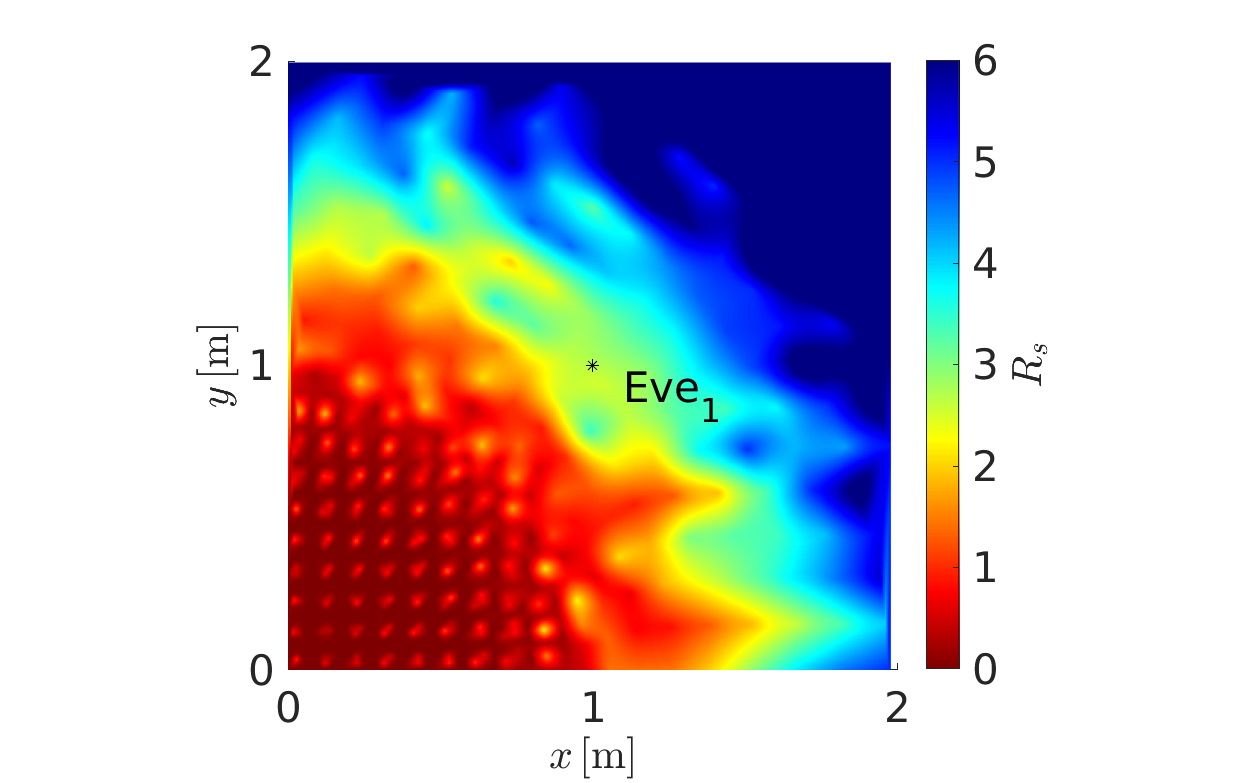}};
			\node[label={[text=white]below left: \LRPSER{} }] at (picSER.north east) {};
			
			\node[below = of picLRAPE,inner sep=0] (picCB){\includegraphics[height=\mapHeightS,trim= {65 0 162 15}, clip]
				{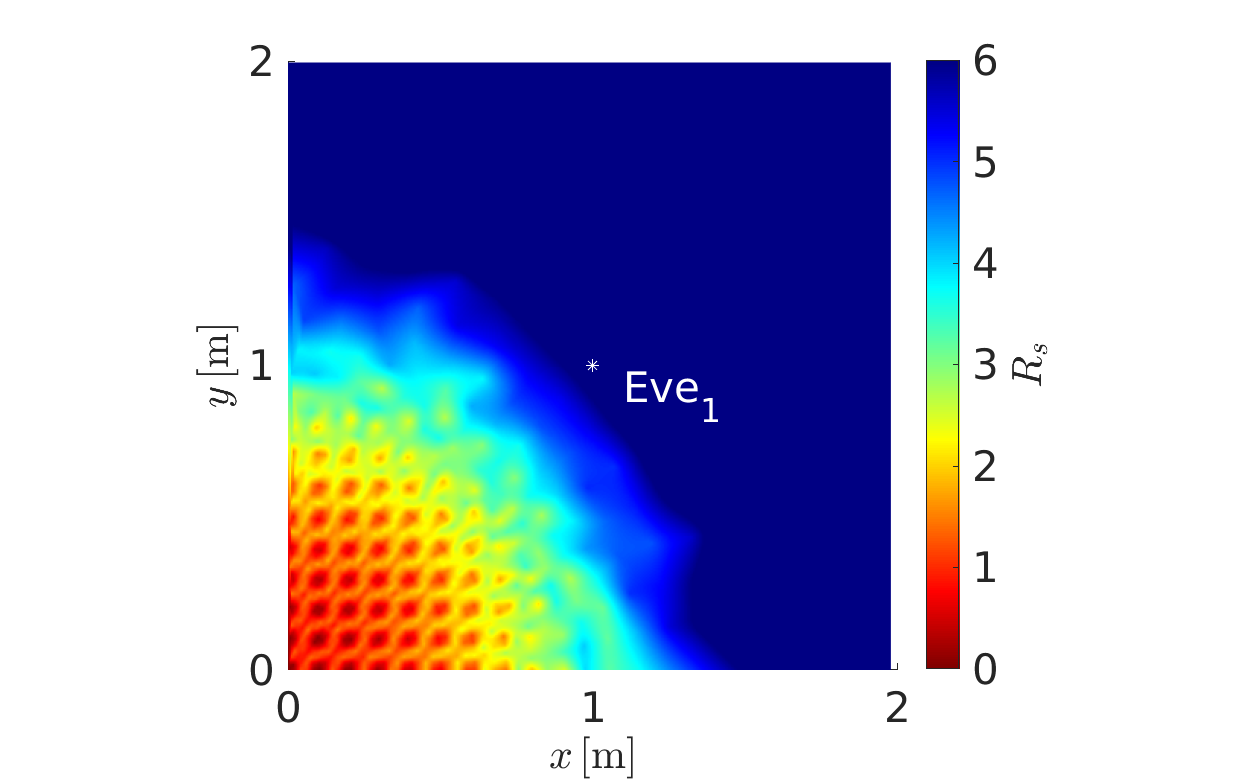}};
			\node[label={[text=white]below left:  \LRPC{} }] at (picCB.north east) {};
			
			\node[right = of picCB,inner sep=0, node distance = 0.5cm] (picCs){\includegraphics[height=\mapHeightS,trim= {65 0 60 15}, clip]
				{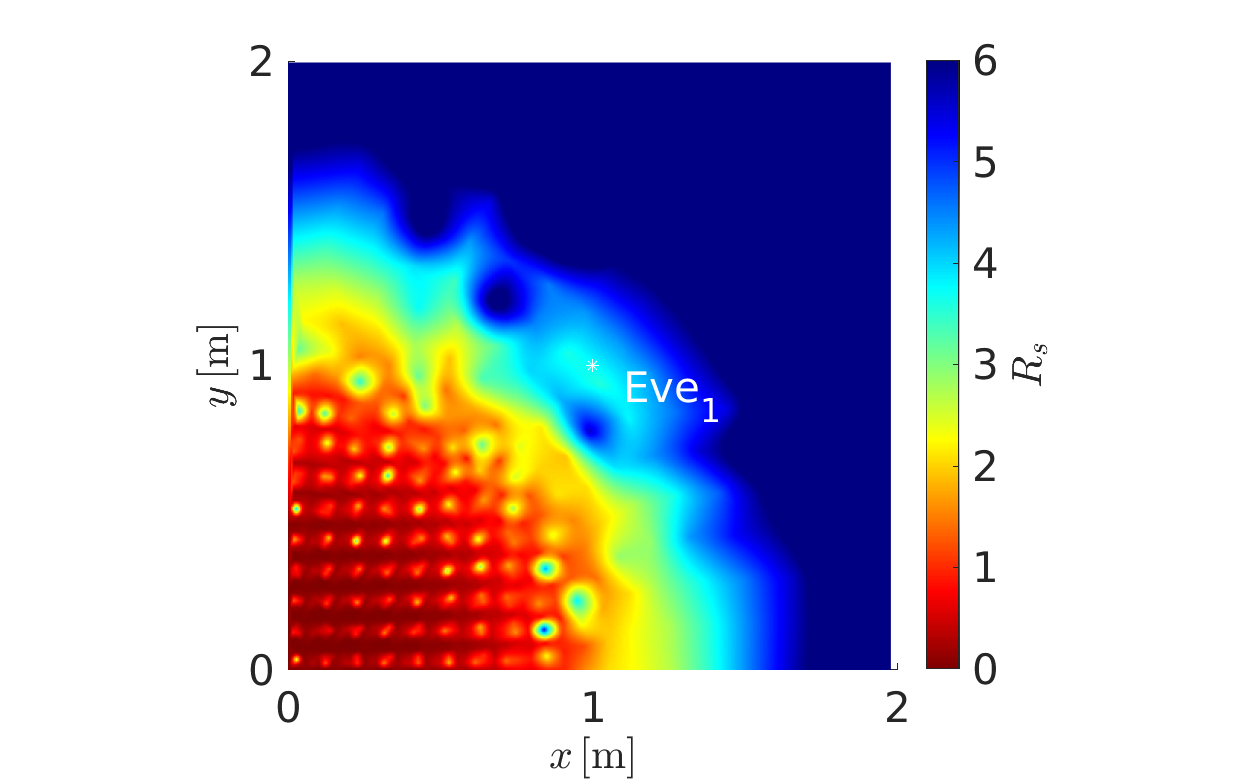}};
			
			\node[right = of picCB,inner sep=0] (picCsinv){\includegraphics[height=\mapHeightS,trim= {65 0 162 15}, clip]
				{135_h3_Atri_m3_G15_B1_2_R500_G15_E1084_2_R20_G15_B3_nF1_19dB_sims1_n20000_CsOptLRAGaussGaussIMin10Matlab_Cs_BGauss_EGauss_Mean_minERot_LE}};
			\node[label={[text=white]below left: \ZCsPE{} }] at (picCsinv.north east) {};
			\end{tikzpicture}
			
			\caption{Comparison of different preprocessing algorithms with respect to the secrecy rate. \label{fig_CsMaps}}
		\end{figure}
	
	In Fig.~\ref{fig_CsMaps} the corresponding secrecy rates are displayed for \LRAPE{}, \LRPSER{}, \LRPC{}, and \ZCsPE{}. When comparing these maps to the SER-Maps, it can be seen that there is a great similarity to the SER-maps. The positions that experience a low SER, also correspond to a low secrecy rate and vice versa. The shape of the insecure areas, where security cannot be guaranteed show the same size for both maps. Consequently, equivalent conclusions may be drawn from both metrics. 
%
%

\section{Conclusion}
	\label{sec_conclusion}
	\noindent
	In this work, the usage of preprocessing for physical-layer security in a wireless MIMO THz-communication scenario has been considered. In this work, preprocessing that is only based on the legitimate receiver has been compared to preprocessing which includes the legitimate receiver and the eavesdropper. For both variants, a rate-based approach has been presented as well as a technique that considered the error performance. Both variants that include the eavesdropper in the optimization achieve a higher  secrecy rate and a lower performance of the eavesdropper. However, there is also a loss in performance for Bob. If the target performance of Bob is fixed, these variants show a larger area where security is not guaranteed.
	
	For each preprocessing variant, a linear and an LRA version have been included. Employing the lattice-based variant enables an improvement in the performance of Bob, without causing a deterioration of the security.


\begin{thebibliography}{10}
	\providecommand{\url}[1]{#1}
	\csname url@samestyle\endcsname
	\providecommand{\newblock}{\relax}
	\providecommand{\bibinfo}[2]{#2}
	\providecommand{\BIBentrySTDinterwordspacing}{\spaceskip=0pt\relax}
	\providecommand{\BIBentryALTinterwordstretchfactor}{4}
	\providecommand{\BIBentryALTinterwordspacing}{\spaceskip=\fontdimen2\font plus
		\BIBentryALTinterwordstretchfactor\fontdimen3\font minus
		\fontdimen4\font\relax}
	\providecommand{\BIBforeignlanguage}[2]{{%
			\expandafter\ifx\csname l@#1\endcsname\relax
			\typeout{** WARNING: IEEEtranS.bst: No hyphenation pattern has been}%
			\typeout{** loaded for the language `#1'. Using the pattern for}%
			\typeout{** the default language instead.}%
			\else
			\language=\csname l@#1\endcsname
			\fi
			#2}}
	\providecommand{\BIBdecl}{\relax}
	\BIBdecl
	
	\bibitem{agrell2002ClosestPointSearch}
	E.~Agrell, T.~Eriksson, A.~Vardy, and K.~Zeger, ``Closest {{Point Search}} in
	{{Lattices}},'' \emph{IEEE Transactions on Information Theory}, vol.~48,
	no.~8, pp. 2201--2214, Aug. 2002.
	
	\bibitem{anderson2006DigitalTransmissionEngineering}
	J.~B. Anderson, \emph{Digital {{Transmission Engineering}}}.\hskip 1em plus
	0.5em minus 0.4em\relax John Wiley \& Sons, Feb. 2006.
	
	\bibitem{bloch2011PhysicalLayerSecurity}
	M.~Bloch and J.~Barros, \emph{Physical--{{Layer Security}}: {{From Information
				Theory}} to {{Security Engineering}}}.\hskip 1em plus 0.5em minus 0.4em\relax
	Cambridge University Press, 2011.
	
	\bibitem{collin1985AntennasRadiowavePropagation}
	R.~E. Collin, \emph{Antennas and {{Radiowave Propagation}}}, 4th~ed.\hskip 1em
	plus 0.5em minus 0.4em\relax New York: McGraw-Hill Inc.,US, Feb. 1985.
	
	\bibitem{fischer2002PrecodingSignalShaping}
	R.~F.~H. Fischer, \emph{Precoding and {{Signal Shaping}} for {{Digital
				Transmission}}}.\hskip 1em plus 0.5em minus 0.4em\relax New York: J.
	Wiley-Interscience, 2002.
	
	\bibitem{fischer2019LatticereductionaidedIntegerforcingEqualization}
	R.~F.~H. Fischer, S.~Stern, and J.~B. Huber, ``Lattice-{{Reduction-Aided}} and
	{{Integer-Forcing Equalization}},'' \emph{Foundations and Trends in
		Communications and Information Theory}, vol.~16, no. 1-2, pp. 1--159, Dec.
	2019.
	
	\bibitem{friis1946NoteSimpleTransmission}
	H.~Friis, ``A {{Note}} on a {{Simple Transmission Formula}},''
	\emph{Proceedings of the IRE}, vol.~34, no.~5, pp. 254--256, May 1946.
	
	\bibitem{giordani20206GNetworksUse}
	M.~Giordani, M.~Polese, M.~Mezzavilla, S.~Rangan, and M.~Zorzi, ``Toward {{6G
			Networks}}: {{Use Cases}} and {{Technologies}},'' \emph{IEEE Communications
		Magazine}, vol.~58, no.~3, pp. 55--61, Mar. 2020.
	
	\bibitem{han2014MultiRayChannelModeling}
	C.~Han, A.~O. Bicen, and I.~F. Akyildiz, ``Multi-{{Ray Channel Modeling}} and
	{{Wideband Characterization}} for {{Wireless Communications}} in the
	{{Terahertz Band}},'' \emph{IEEE Transactions on Wireless Communications},
	vol.~14, no.~5, pp. 2402--2412, May 2014.
	
	\bibitem{khisti2010SecureTransmissionMultiple}
	A.~Khisti and G.~W. Wornell, ``Secure {{Transmission With Multiple
			Antennas}}---{{Part II}}: {{The MIMOME Wiretap Channel}},'' \emph{IEEE
		Transactions on Information Theory}, vol.~56, no.~11, pp. 5515--5532, Nov.
	2010.
	
	\bibitem{klinc2011LDPCCodesGaussian}
	D.~Klinc, J.~Ha, S.~W. McLaughlin, J.~Barros, and B.-J. Kwak, ``{{LDPC Codes}}
	for the {{Gaussian Wiretap Channel}},'' \emph{IEEE Transactions on
		Information Forensics and Security}, vol.~6, no.~3, pp. 532--540, Sep. 2011.
	
	\bibitem{leung-yan-cheong1978GaussianWireTapChannel}
	S.~{Leung-Yan-Cheong} and M.~Hellman, ``The {{Gaussian Wire-Tap Channel}},''
	\emph{IEEE Transactions on Information Theory}, vol.~24, no.~4, pp. 451--456,
	Jul. 1978.
	
	\bibitem{li2013AlternatingOptimizationAlgorithm}
	Q.~Li, M.~Hong, H.-T. Wai, W.-K. Ma, Y.-F. Liu, and Z.-Q. Luo, ``An alternating
	optimization algorithm for the {{MIMO}} secrecy capacity problem under sum
	power and per-antenna power constraints,'' in \emph{{{IEEE International
				Conference}} on {{Acoustics}}, {{Speech}} and {{Signal Processing}}}, May
	2013, pp. 4359--4363.
	
	\bibitem{loyka2015AlgorithmGlobalMaximization}
	S.~Loyka and C.~D. Charalambous, ``An {{Algorithm}} for {{Global Maximization}}
	of {{Secrecy Rates}} in {{Gaussian MIMO Wiretap Channels}},'' \emph{IEEE
		Transactions on Communications}, vol.~63, no.~6, pp. 2288--2299, Jun. 2015.
	
	\bibitem{mukherjee2009UtilityBeamformingStrategies}
	A.~Mukherjee and A.~L. Swindlehurst, ``Utility of beamforming strategies for
	secrecy in multiuser {{MIMO}} wiretap channels,'' in \emph{47th {{Annual
				Allerton Conference}} on {{Communication}}, {{Control}}, and {{Computing}}
		({{Allerton}})}.\hskip 1em plus 0.5em minus 0.4em\relax Monticello, IL, USA:
	IEEE, Sep. 2009, pp. 1134--1141.
	
	\bibitem{mukherjee2011RobustBeamformingSecurity}
	------, ``Robust {{Beamforming}} for {{Security}} in {{MIMO Wiretap Channels
			With Imperfect CSI}},'' \emph{IEEE Transactions on Signal Processing},
	vol.~59, no.~1, pp. 351--361, Jan. 2011.
	
	\bibitem{mukherjee2021SecrecyCapacityMIMO}
	A.~Mukherjee, B.~Ottersten, and L.-N. Tran, ``On the {{Secrecy Capacity}} of
	{{MIMO Wiretap Channels}}: {{Convex Reformulation}} and {{Efficient Numerical
			Methods}},'' \emph{IEEE Transactions on Communications}, vol.~69, no.~10, pp.
	6865--6878, Oct. 2021.
	
	\bibitem{nguyen2020LowComplexityAlgorithmAchieving}
	T.~V. Nguyen, Q.-D. Vu, M.~Juntti, and L.-N. Tran, ``A {{Low-Complexity
			Algorithm}} for {{Achieving Secrecy Capacity}} in {{MIMO Wiretap
			Channels}},'' in \emph{{{IEEE International Conference}} on
		{{Communications}} ({{ICC}})}, Jun. 2020, pp. 1--6.
	
	\bibitem{oggier2011SecrecyCapacityMIMO}
	F.~Oggier and B.~Hassibi, ``The {{Secrecy Capacity}} of the {{MIMO Wiretap
			Channel}},'' \emph{IEEE Transactions on Information Theory}, vol.~57, no.~8,
	pp. 4961--4972, Aug. 2011.
	
	\bibitem{piesiewicz2007ScatteringAnalysisModeling}
	R.~Piesiewicz, C.~Jansen, D.~Mittleman, T.~{Kleine-Ostmann}, M.~Koch, and
	T.~K{\"u}rner, ``Scattering {{Analysis}} for the {{Modeling}} of {{THz
			Communication Systems}},'' \emph{IEEE Transactions on Antennas and
		Propagation}, vol.~55, no.~11, pp. 3002--3009, Nov. 2007.
	
	\bibitem{rappaport2019WirelessCommunicationsApplications}
	T.~S. Rappaport, Y.~Xing, O.~Kanhere, S.~Ju, A.~Madanayake, S.~Mandal,
	A.~Alkhateeb, and G.~C. Trichopoulos, ``Wireless {{Communications}} and
	{{Applications Above}} 100 {{GHz}}: {{Opportunities}} and {{Challenges}} for
	{{6G}} and {{Beyond}},'' \emph{IEEE Access}, vol.~7, pp. 78\,729--78\,757,
	2019.
	
	\bibitem{schraml2021MultiuserMIMOConcept}
	M.~G. Schraml, R.~T. Schwarz, and A.~Knopp, ``Multiuser {{MIMO Concept}} for
	{{Physical Layer Security}} in {{Multibeam Satellite Systems}},'' \emph{IEEE
		Transactions on Information Forensics and Security}, vol.~16, pp. 1670--1680,
	2021.
	
	\bibitem{schulz2023LatticeReductionAidedPreequalizationPhysicalLayer}
	R.~Schulz and R.~F.~H. Fischer, ``Lattice-{{Reduction-Aided Preequalization}}
	for {{Physical-Layer Security}} in {{Wireless THz-Communication}},'' in
	\emph{26th {{International ITG Workshop}} on {{Smart Antennas}} and 13th
		{{Conference}} on {{Systems}}, {{Communications}}, and {{Coding}}}, Feb.
	2023.
	
	\bibitem{schulz2024LatticeReductionAidedPreprocessingPhysicalLayer}
	------, ``Lattice-{{Reduction-Aided Preprocessing}} for {{Physical-Layer
			Security}},'' in \emph{19th {{International Symposium}} on {{Wireless
				Communication Systems}} ({{ISWCS}})}, Rio de Janeiro, Jul. 2024, in Press.
	
	\bibitem{stern2017OptimalFactorizationLatticeReductionAided}
	S.~Stern and R.~F.~H. Fischer, ``Optimal {{Factorization}} in
	{{Lattice-Reduction-Aided}} and {{Integer-Forcing Linear Equalization}},'' in
	\emph{11th {{International ITG Conference}} on {{Systems}},
		{{Communications}} and {{Coding}}}, Feb. 2017.
	
	\bibitem{utkovski2019LearningRadioMaps}
	Z.~Utkovski, P.~Agostini, M.~Frey, I.~Bjelakovic, and S.~Stanczak, ``Learning
	{{Radio Maps}} for {{Physical-Layer Security}} in the {{Radio Access}},'' in
	\emph{{{IEEE}} 20th {{International Workshop}} on {{Signal Processing
				Advances}} in {{Wireless Communications}} ({{SPAWC}})}.\hskip 1em plus 0.5em
	minus 0.4em\relax Cannes, France: IEEE, Jul. 2019, pp. 1--5.
	
	\bibitem{viswanath2003SumCapacityVector}
	P.~Viswanath and D.~Tse, ``Sum capacity of the vector {{Gaussian}} broadcast
	channel and uplink--downlink duality,'' \emph{IEEE Transactions on
		Information Theory}, vol.~49, no.~8, pp. 1912--1921, Aug. 2003.
	
	\bibitem{weingarten2006CapacityRegionGaussian}
	H.~Weingarten, Y.~Steinberg, and S.~Shamai, ``The {{Capacity Region}} of the
	{{Gaussian Multiple-Input Multiple-Output Broadcast Channel}},'' \emph{IEEE
		Transactions on Information Theory}, vol.~52, no.~9, pp. 3936--3964, Sep.
	2006.
	
	\bibitem{wyner1975WireTapChannel}
	A.~D. Wyner, ``The {{Wire-Tap Channel}},'' \emph{Bell System Technical
		Journal}, vol.~54, no.~8, pp. 1355--1387, 1975.
	
	\bibitem{zamir2014LatticeCodingSignals}
	R.~Zamir, \emph{Lattice {{Coding}} for {{Signals}} and {{Networks}}: {{A
				Structured Coding Approach}} to {{Quantization}}, {{Modulation}}, and
		{{Multiuser Information Theory}}}.\hskip 1em plus 0.5em minus 0.4em\relax
	Cambridge University Press, Aug. 2014.
	
	\bibitem{zhao2015RobustBeamformingDesign}
	P.~Zhao, M.~Zhang, H.~Yu, H.~Luo, and W.~Chen, ``Robust {{Beamforming Design}}
	for {{Sum Secrecy Rate Optimization}} in {{MU-MISO Networks}},'' \emph{IEEE
		Transactions on Information Forensics and Security}, vol.~10, no.~9, pp.
	1812--1823, Sep. 2015.
	
\end{thebibliography}

\end{document}